\documentclass[12pt]{iopart}

\usepackage{booktabs}
\usepackage{colortbl}  
\usepackage{xcolor}    
\usepackage{caption}
\usepackage{graphicx}
\usepackage{biblatex}  
\usepackage[normalem]{ulem}  

\definecolor{lightgray}{gray}{0.9}
\definecolor{lightblue}{rgb}{0.8, 0.85, 1}
\definecolor{saddlebrown}{RGB}{139,69,19}
\definecolor{softred}{RGB}{220, 80, 80}

\begin{document}

\title[]{Impurity peaking of SPARC H-modes: a sensitivity study on physics and engineering assumptions.}

\author{M. Muraca$^1$, P. Rodriguez-Fernandez$^1$, J. Hall$^1$, N.T. Howard$^1$, D. Fajardo$^2$, G. Tardini$^2$, C. F. B. Zimmermann$^3$, T. Body$^4$}

\address{$^1$MIT Plasma Science and Fusion Center, 167 Albany St, Cambridge, MA 02139, USA \\
$^2$Max Planck Institute for Plasma Physics, Boltzmannstrasse 2, Garching bei M\"unchen, 85748, Germany \\
$^3$Department of Applied Physics and Applied Mathematics, Columbia University, New York 10027, USA \\
$^4$Commonwealth Fusion Systems, 117 Hospital Rd, Devens, MA 01434, USA}
\ead{mmuraca@mit.edu}
\vspace{10pt}
\begin{indented}
\item[]December 2025
\end{indented}

\begin{abstract}
In this paper, an overview of the impurity transport for three H-mode plasmas in the upcoming SPARC tokamak has been provided. The simulations have been performed within the ASTRA+STRAHL framework, using FACIT and TGLF-SAT2 to predict, respectively, neoclassical and turbulent core transport, while a neural network trained on EPED simulations has been employed to calculate the pedestal height and width self-consistently. A benchmark with previous simulations at constant impurity fraction has been provided for three H-modes, spanning different plasma current and magnetic field values. For a scenario, additional simulations have been performed to account for uncertainties in the modeling assumptions. The predictions are nearly insensitive to changes in the top of pedestal W concentrations. Varying the Ar pedestal concentration has shown a small effect on the impurity peaking and nearly constant fusion gain values, due to multiple effects on pedestal pressure, main ion dilution and density peaking. The inclusion of rotation in ASTRA simulations has shown minimal impact on confinement and impurity transport predictions. An exploratory study has been provided with a first set of simulations treating D and T separately, experiencing a maximum fusion power at 55-45\% DT fuel composition, and an asymmetric distribution with respect to the D concentration. All the results, including sensitivity scans of toroidal velocity and ion temperature and density gradients, highlighted that turbulent impurity transport prevails on the neoclassical component, aligning with previous ITER predictions, and suggesting that next generation devices like SPARC, operating at low collisionality, will experience low W accumulation.
\end{abstract}

\section{Introduction}
The prediction of transport and confinement is crucial to determine the fusion performance of next generation devices, like ITER \cite{loarte_new_2025}, DEMO \cite{hong_review_2022}, SPARC \cite{creely_overview_2020} and ARC \cite{sorbom_arc_2015, kuang_conceptual_2018}. Such a task can be performed with different strategies which vary depending on the precision and the computational time required by the models employed. Typically, models which can be run in real-time but have a limited amount of physics ingredients are referred to as \textit{low-fidelity}. Examples are scaling laws \cite{transport_chapter_1999} or data-driven models \cite{meneghini_neural-network_2020, ho_neural_2021}, which are derived from experimental observations on existing machines, fitting procedures and statistical regressions. While such models are very useful to identify general trends and scope the design of future devices, their use for quantitative predictions should be approached with caution. On the other hand, \textit{high-fidelity} simulations are usually conducted with non-linear gyrokinetic (GK) codes \cite{jenko_massively_2000, peeters_effect_2004, kotschenreuther_comparison_1995, candy_eulerian_2003, grandgirard_drift-kinetic_2006}, which are capable of reproducing the physics accurately, but they remain computationally expensive, requiring up to several months to converge. These models are crucial to identify interesting physics and explore sophisticated effects like multi-scale turbulent interactions and saturation mechanisms, but their usage is limited by existing computational resources, and they can not be used to obtain extensive databases. The \textit{medium-fidelity} simulations represent a good compromise between good predictive capability and low computational expenses. Example of medium-fidelity approaches are quasi-linear transport models like TGLF \cite{staebler_theory-based_2007, staebler_verification_2021} and QualiKiZ \cite{bourdelle_new_2007}, whose fluxes can be found through a linear eigenvalue solver and a saturation rule previously fitted on non-linear GK simulations. These models typically require up to a few minutes to run. \\
Generally, transport models can be run at specific radial locations to investigate the dominant instability, or in an integrated way, connecting a series of radial points from the magnetic axis to the separatrix. The second approach, called "integrated modeling", evolves self-consistently the power and kinetic profiles, allowing to predict the fusion performance and confinement of future machines. Examples of integrated modeling codes are ASTRA \cite{pereverzev_astra_1991, pereverzev_astra_2002}, TRANSP \cite{pankin_transp_2025} or JINTRAC \cite{romanelli_jintrac_2014}. These codes are also referred to as flux-matching transport solvers, since they evolve two fluxes terms (the transported flux and the algebraic sum of sources and sinks) until they converge to the steady-state profiles. \\
The integrated transport modeling has been historically mainly conducted with medium fidelity models, but recent progress has opened up the possibility to employ non-linear GK models \cite{di_siena_global_2022, rodriguez-fernandez_nonlinear_2022, rodriguez-fernandez_enhancing_2024, guttenfelder_predictions_2025}. Within the integrated modeling framework several assumptions can be made on the uncertain parameters, without including a detailed model to predict them. This allows to speed up the simulations and generate extensive databases. Examples of uncertain parameters assumed in the simulations are the H-mode pedestal density, fuel mix concentration or impurity radial profiles. It has been demonstrated in \cite{muraca_integrated_2025} that such parameters can dramatically impact the fusion gain and power. In particular the W concentration has been found to be a crucial parameter which governs the radiation and the H-mode accessibility. This has motivated a study of the impurity transport in the core, which was previously prescribed through fixed concentrations, as in the SPARC Physics Basis \cite{creely_overview_2020} and followup work. Such modeling aims at validating previous methodologies employed to quantify the variations of performance, and explore the physics governing the impurity peaking in the core. Therefore, in this paper we predicted the impurity transport for 3 SPARC H-mode scenarios that are being considered for operation, covering different plasma current and magnetic field values. Scans of input power and pedestal density have been performed to study the robustness of the results. A previously unexplored H-mode scenario has been selected to perform additional scans of the boundary conditions for Ar and W concentrations, rotation and DT fuel mix composition. These sensitivity studies accounted for uncertainties on input parameters, showing minimal deviations of impurity peaking and fusion gain. \\
The remainder of the paper is organized as follows: in section 2, the framework of the simulations is described; in section 3, the SPARC H-mode scenarios are introduced and the nominal simulation results are shown, including a benchmark with previous results; in section 4 the Ar and W concentrations at the pedestal top are varied and the results are discussed; in section 5 the impact of rotation on impurity transport is explored; in section 6 a variation of DT fuel mix composition is performed and the results are analyzed; in section 7, the article is summarized and the conclusions are discussed, including an outlook for future work.

\section{Simulation Setup}
In order to predict the performance of a discharge, a series of models needs to be integrated in a framework which predicts plasma kinetic profiles. In the present paper, this task has been performed with the time-dependent transport solver ASTRA \cite{pereverzev_astra_1991}. Although ASTRA can simulate the time trajectories of a plasma pulse, here we focused on the stationary flat-top phase of three H-mode scenarios, where equilibrium, kinetic profiles and fusion power have reached a converged state. The plasma equilibrium has been calculated with SPIDER \cite{ivanov_new_2005}, which performs fixed-boundary computations. The last closed flux surface (LCFS) coordinates are obtained with FreeGS \cite{noauthor_freegs-plasmafreegs_2024} simulations. Both SPIDER and FreeGS solve the Grad-Shafranov equation \cite{grad_hydromagnetic_1958, shafranov_plasma_1966}. \\
Core turbulent transport in modern, high-power tokamaks is driven by micro-instabilities like Electron Temperature Gradient (ETG), Trapped Electron Mode (TEM) and Ion Temperature Gradient (ITG) \cite{weiland_collective_2000, dimits_comparisons_2000, garbet_introduction_2006}. Here, turbulent transport in the core is predicted using the quasi-linear TGLF model \cite{staebler_verification_2021}. TGLF is a common tool used in reactor design studies, which has been widely validated on different machines and plasma scenarios, showing reliability \cite{rodriguez-fernandez_predict-first_2019, rodriguez-fernandez_perturbative_2019, angioni_confinement_2022, baiocchi_turbulent_2015, creely_validation_2017, staebler_quasilinear_2024}. The control settings adopted in this project include electromagnetic effects ($\delta A_\parallel$), five plasma species, Miller geometry \cite{stacey_representation_2009}, saturation rule SAT2 \cite{staebler_verification_2021} and a maximum of 6 parallel basis functions.\\
The neoclassical impurity transport has been calculated with FACIT \cite{fajardo_analytical_2022, fajardo_analytical_2023}, an analytical model which includes the effect of rotation and poloidal asymmetries, providing flux-surface averaged coefficients for arbitrary values of collisionality, charge, mass and radial position in the confined region. The atomic processes of impurities (e.g. ionization and recombination) are calculated self-consistently in STRAHL \cite{dux_strahl_2006}, which is called within ASTRA at every computation time step. STRAHL calculates also the radiated power and transport of impurities for every charge state, taking as input the wall source, diffusivity ($D$) and convection ($v$) for every impurity involved in the simulations. While FACIT can provide $D$ and $v$ separately, TGLF calculates directly the combined particle flux. Thereofore, in order to separate D and v, TGLF is called 2 times at every time step, once with a different impurity density gradient and once with the actual density profile, similarly as in \cite{fajardo_full-radius_2024}. This allows to separate the convective from the diffusive term, under the assumption of impurities being trace species, which do not impact the background turbulence.\\
The plasmas considered in this study are H-modes, therefore the pedestal stability must be evaluated in the context of peeling-ballooning theory. It is worth to mention that ELMy H-mode regimes are operationally challenging because of their high power and particle fluxes on Plasma-Facing Components (PFCs). However, the evaluation of operational strategies to mitigate ELMs is beyond the scope of the paper, and we consider such regimes as upper performance limits. The prediction of pedestal pressure is done using a neural network (NN) trained on EPED simulations of SPARC \cite{snyder_first-principles_2011}, following a procedure similar to \cite{meneghini_self-consistent_2017}. This model was already introduced in \cite{muraca_integrated_2025}, where its details are described. This model self-consistently updates the pedestal height as plasma gradients and stored energy evolve, allowing to correctly capture the effect of global beta on the pedestal pressure. In this NN framework, the pedestal density is an input parameter, as it depends on fueling, neutral penetration, and edge physics, which are not modeled here. The turbulent transport of impurities in the edge is tuned to achieve target concentrations at the top of pedestal. This choice is imposed by the lack of robust models which can be employed in an integrated framework. This represents a lack of self-consistency, but the task of this paper is to quantify core impurity peaking. Therefore, in this context, the pedestal model should only provide reasonable pedestal boundary conditions, while the precise quantification of edge transport is beyond the scope of the present work and the capabilities of this framework. However, a sensitivity study of pedestal concentrations has been performed in the paper to account for these uncertainties. \\
SPARC auxiliary power will be provided by the Ion Cyclotron Resonance Heating (ICRH) \cite{lin_physics_2020}. Since there is no self-consistent model to reproduce the ICRH absorption profiles within ASTRA, simulations were conducted using TRANSP \cite{pankin_transp_2024}, coupled with TORIC \cite{brambilla_numerical_1999} and FPPMOD \cite{hammett_fast_1986}, to model ICRF wave propagation and Fokker Planck collisions. This is consistent with the methodology adopted in \cite{muraca_integrated_2025}, ensuring that any variation in the results can be attributed to different assumptions on the impurity modeling, while keeping the ICRH input power constant. The resulting ICRH deposition profiles are imported into ASTRA simulations. The ohmic power is $P_{OH}=\sigma E_z^2$, where $\sigma$ is the plasma conductivity and $E_z$ is the electric field in the toroidal direction. The collisional exchange is $P_{ex}=0.00246 \Lambda n_e n_i Z^2(T_e-T_i)(A T_e^{3/2})^{-1}$, where $\Lambda$ is the Coloumb logarithm, $n_e$ and $n_i$ are the electron and ion densities, $T_e$ and $T_i$ are the electron and ion temperatures, $Z$ and $A$ are the charge and mass of the main ion species. The fusion power is $P_{fus}=5 P_{\alpha} = 5\cdot5.632 n_D n_T \sigma_{DT}$, with an $f_{ion}$ which calculates the fraction of power exchanged with ions and electrons, calculated as in \cite{barnett_plasmas_2013}. As previously mentioned, $P_{rad}$ is computed self-consistently in STRAHL, considering Bremsstrahlung, synchrotron and impurity line radiations for every charge state. Given the uncertainties in the neutral source penetration, especially in the typically stiff pedestals of H-modes, and the lack of reliable pedestal transport models, a top of pedestal density is assumed as boundary condition, and no D or T core particle source has been used in the simulations.\\
SPARC's compact design and the consequently high fluxes require a careful management of PFCs loads, to mitigate technological damages. Detachment, which represents a possible strategy to protect the divertor \cite{kallenbach_impurity_2013, soukhanovskii_review_2017, krasheninnikov_physics_2017}, is often achieved by injecting impurities into the Scrape-Off Layer (SOL), e.g. Ar, Ne or N. A small fraction of such impurities penetrates into the confined region, affecting the performance via radiation and dilution of the main ions. The presence of W, introduced by the erosion of PFCs, as well as plasma minority ions, used to efficiently heat the plasma with ICRH \cite{van_eester_minority_2012}, further influences the core confinement. Therefore, in this paper, W, Ar and He3/H transport has been simultaneously modeled to quantify the impurity density peakings and their impact on performance. It is worth to mention that He3/H minorities are treated as thermal species, since a self-consistent Fokker-Planck collisional model is missing in ASTRA. This assumption is not valid near-axis, where most of the wave is absorbed, but in this work we are not addressing the effect of fast ions on turbulence, and the modeling of minority species aims at providing a realistic DT dilution and quantifying the He3/H core penetration to determine wether sufficient near-axis concentrations are reached to efficiently absorb ICRH power. \\
Sawteeth are expected to modify the safety factor and kinetic profiles in SPARC plasmas, particularly due to the absence of neutral beams, which could modify the q profile through current drive \cite{rodriguez-fernandez_predictions_2020}. Although this study is not focused on magnetohydrodynamics (MHD), reliably calculating the safety factor profile is essential, since it directly affects transport. Therefore, the Kadomtsev model \cite{kadomtsev_disruptive_1975} has been used within ASTRA to predict the sawtooth inversion radius and flatten the safety factor profile. Additionally, ASTRA requires the sawtooth period as an input to trigger the profile relaxation. This period has been calculated with the Porcelli model \cite{porcelli_model_1996} in TRANSP simulations for the 3 different analyzed scenarios, and it has subsequently been used as fixed parameter in ASTRA. A sensitivity study of the sawtooth period has been performed in \cite{muraca_integrated_2025}, revealing a weak effect on the fusion power, which is calculated right before the sawtooth crash. \\

\section{Benchmark of impurity transport simulations}
In this section we discuss the results of our impurity modeling for three SPARC H-modes, and we compare them with simulations performed assuming fixed radial concentrations. The H-modes analyzed are the Primary Reference Discharge (PRD), the Reduced Magnetic Field scenario (H8 or 8T), and the Reduced Plasma Current scenario (H12). The first two H-modes were already introduced in \cite{muraca_integrated_2025}, while the third one was analyzed for the first time in this paper. These pulses are described separately in the remainder of the section.

\subsection{Primary Reference Discharge}
\begin{figure}[h]
    \centering
    \includegraphics[width=0.45\textwidth]{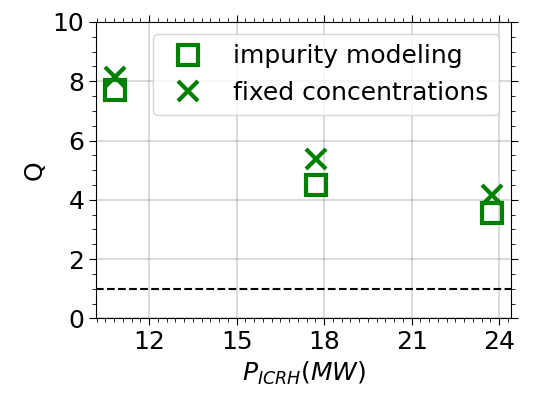}
    \includegraphics[width=0.45\textwidth]{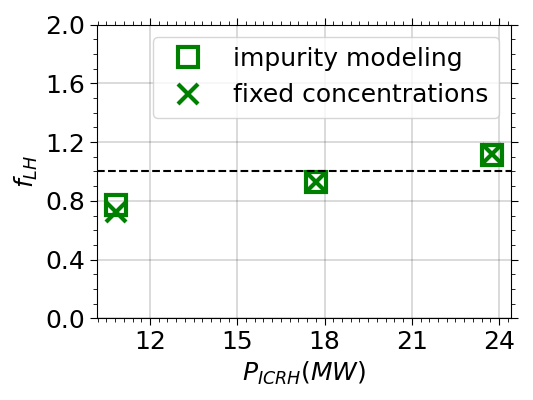}
    \includegraphics[width=0.45\textwidth]{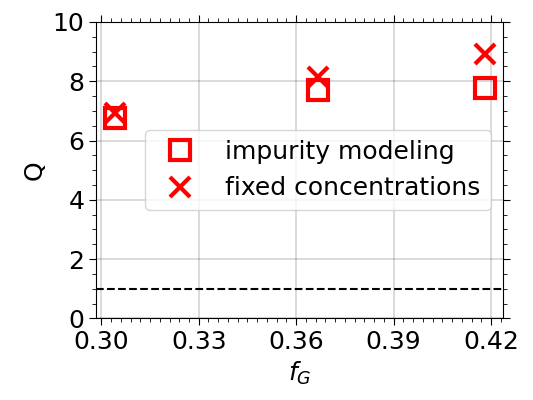}
    \includegraphics[width=0.45\textwidth]{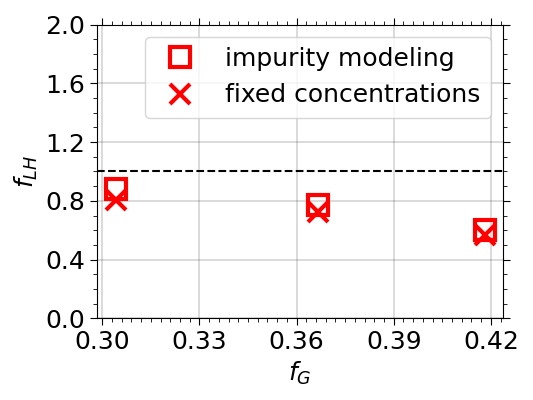}
    \caption{On the left and right are shown respectively fusion gain and $f_{LH}$ values for an ICRH power (top plots, green) and density (bottom plots, red) scan of the PRD scenario. The squares indicate simulations with impurity transport, while the crosses are simulated keeping fixed radial impurity concentrations.}
    \label{fig:PRD_scans}
\end{figure}

\begin{figure}[h]
    \centering
    \includegraphics[width=0.45\textwidth]{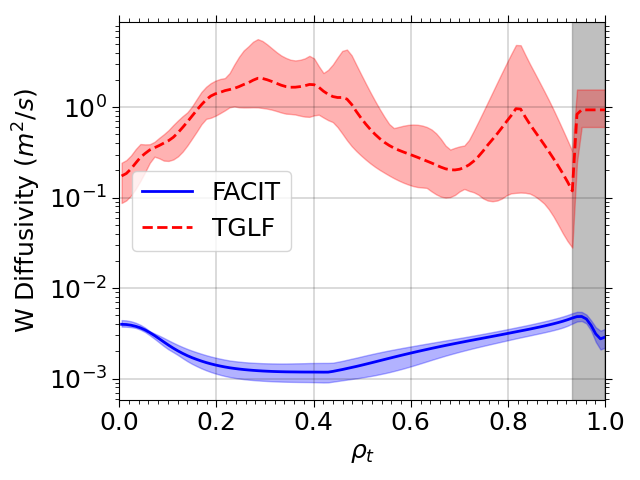}
    \includegraphics[width=0.45\textwidth]{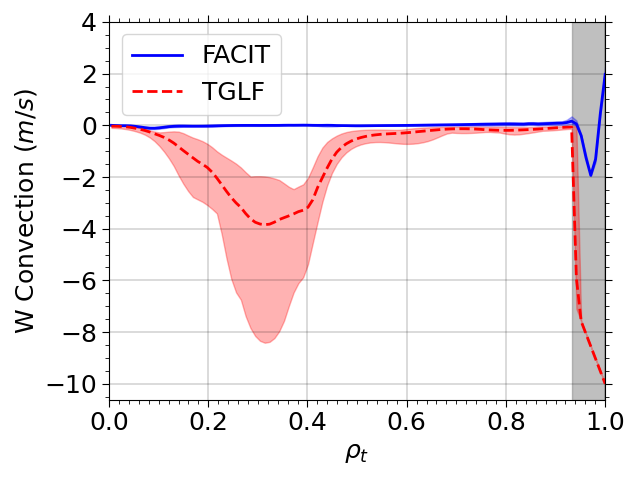}
    \includegraphics[width=0.45\textwidth]{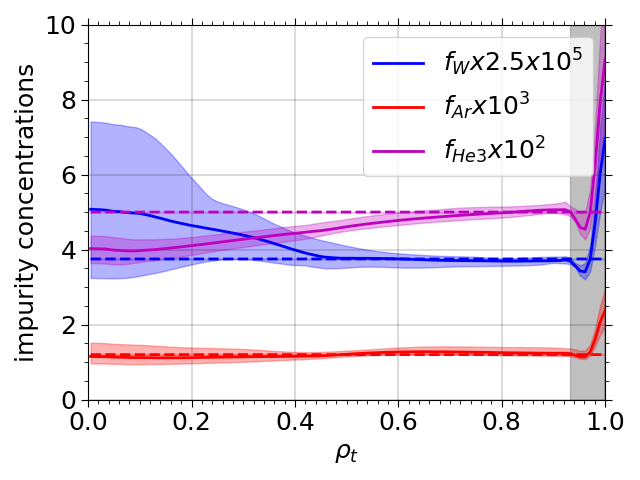}
    \includegraphics[width=0.45\textwidth]{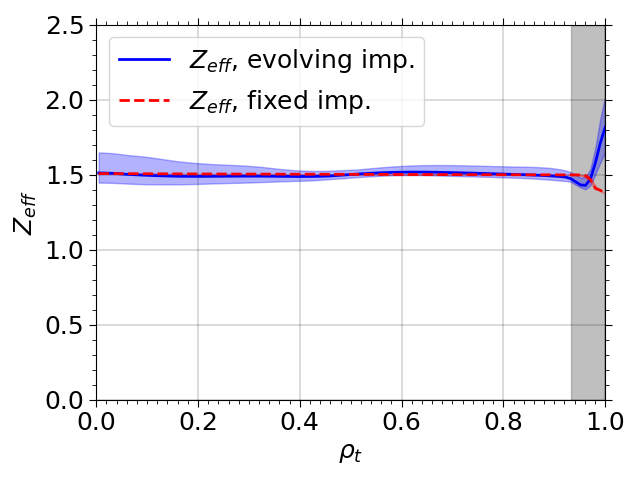}
    \caption{PRD profiles. Top left: turbulent (red) and neoclassical (blue) W diffusivity; top right: turbulent (red) and neoclassical (blue) W convection; bottom left: impurity concentrations; bottom right: $Z_{eff}$. In the bottom plots, the dashed lines indicate the simulations with fixed concentrations. The colored shades indicate the maximum and minimum values computed across the scans. The gray area is from top of pedestal to the separatrix.}
    \label{fig:PRD_imp_profiles}
\end{figure}

The PRD is a scenario envisioned in SPARC to maximize the reachable fusion gain. The main parameters are listed in table \ref{tab:PRDvalues}.
\begin{table}[h]
  \centering
  \caption{List of the main global and engineering parameters of the SPARC Primary Reference Discharge. Their definition is in \cite{creely_overview_2020}.}
  \begin{tabular}{cc}
    \toprule
    \rowcolor{lightblue} 
    \textbf{Parameter} & \textbf{Value} \\
    \midrule
          $B_t$  &  12.2 T  \\
          $I_p$  &  8.7 MA  \\
          $R_0$  &  1.85 m  \\
          $a$  &  0.57 m  \\
          $k_{sep}$  &  1.97  \\
          $\delta_{sep}$  &  0.54  \\
          $P_{ICRH}$  &  11 MW  \\
          $\langle n_e \rangle$  &  $3.1\cdot10^{20}m^{-3}$  \\
          $q^*_{Uckan}$  &  3.05  \\
          $f_G$  &  0.37  \\
          $Z_{eff}$  &  1.5  \\
    \bottomrule
  \end{tabular}
  \label{tab:PRDvalues}
\end{table}
This scenario adopts He3 minority species for efficient ICRH heating. As previously described, W, Ar and He3 sources and edge transport coefficients are artificially set to match top of pedestal target values of $f_{W}$, $f_{He3}$ and $Z_{eff}$ from previous modeling of the same discharge \cite{muraca_integrated_2025}. The same tuning approach has been used for the other scenarios analyzed in this paper. Therefore, given the arbitrariness behind the choice of edge transport and source, we will mainly refer directly to the resulting pedestal concentrations in the remainder of the paper. This choice reflects the scope of the present work, which does not attempt to determine limits on the impurity source originating from the scrape-off layer (SOL), but rather quantifying the core impurity peaking. \\
Simulations with and without impurity transport have been conducted, and the resulting fusion gain and $f_{LH}=P_{sep}/P_{LH}$ values, where $P_{sep}$ and $P_{LH}$ are respectively the power at the separatrix and the LH transition power threshold according to the Schmidtmayr scaling \cite{schmidtmayr_investigation_2018}, are plotted in figure \ref{fig:PRD_scans}.
 Here, the squares are simulations with impurity transport, while the crosses are simulated assuming fixed concentrations. Red and green symbols refer respectively to the density and ICRH power scans. The results show a very good benchmark with almost no deviation, validating the simplified approach of assuming fixed flat concentrations. The core impurity peaking is dominated by the turbulence, while the neoclassical contribution is negligible in these conditions. Such finding is consistent with \cite{fajardo_integrated_2024}, and is justified by high ion temperature gradient and low rotation conditions, which typically exhibit a small neoclassical pinch \cite{angioni_neoclassical_2014, casson_theoretical_2015, fajardo_analytical_2023}.
 The neoclassical and turbulent W convections and diffusivities are compared in figure \ref{fig:PRD_imp_profiles}, showing that D and V are both dominated by the turbulence. The same result has been found for Ar, showing respectively a 3\% and 4\% maximum contribution of neoclassical diffusivity and convection. In the same figure, $f_W$, $f_{He3}$ and $Z_{eff}$  are also shown for the simulations with fixed and evolving impurity densities, indicating that a 4\% He3 on-axis concentration can be reached with a 5\% value at top of pedestal. In figure \ref{fig:PRD_imp_profiles} the solid or dashed lines indicate the mean profiles, while the shaded areas around them show the maximum and minimum values obtained across all the simulations. The same style will be used to show profiles variations in the remainder of the article.

\subsection{Reduced Magnetic Field, $I_p=5.7~MA$}
\begin{figure}[h]
    \centering
    \includegraphics[width=0.45\textwidth]{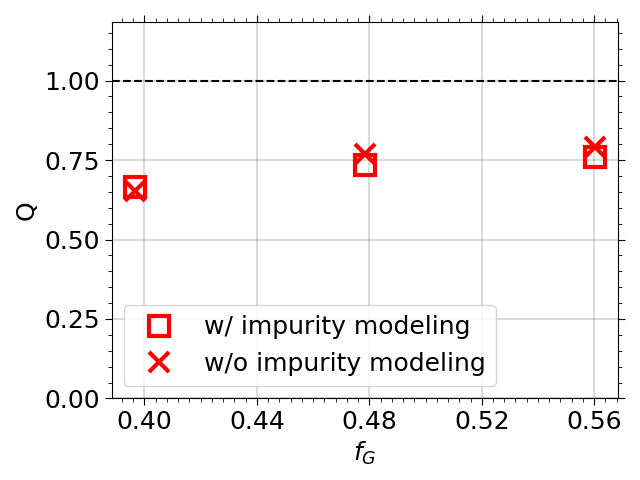}
    \includegraphics[width=0.45\textwidth]{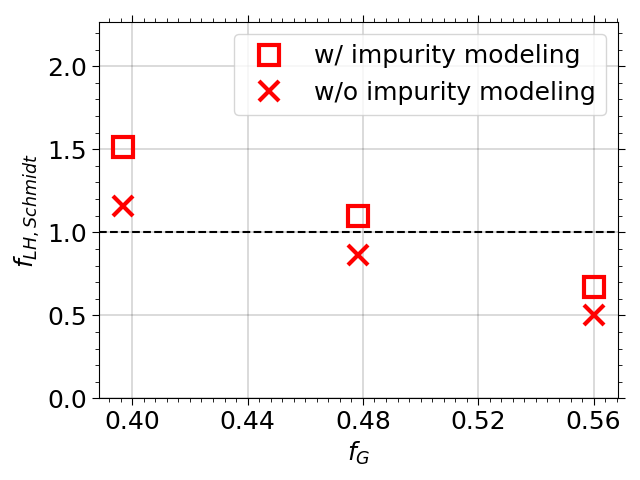}
    \caption{On the left and right are shown respectively fusion gain and $f_{LH}$ values for a density scan of the 8T H-mode scenario, with $P_{ICRH}=25MW$. The squares indicate simulations with impurity transport, while the crosses are simulated keeping fixed radial impurity concentrations.}
    \label{fig:H8_scan}
\end{figure}

\begin{figure}[h]
    \centering
    \includegraphics[width=0.45\textwidth]{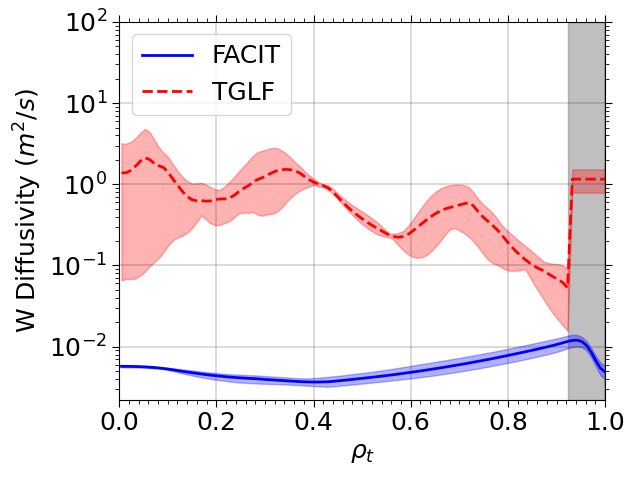}
    \includegraphics[width=0.45\textwidth]{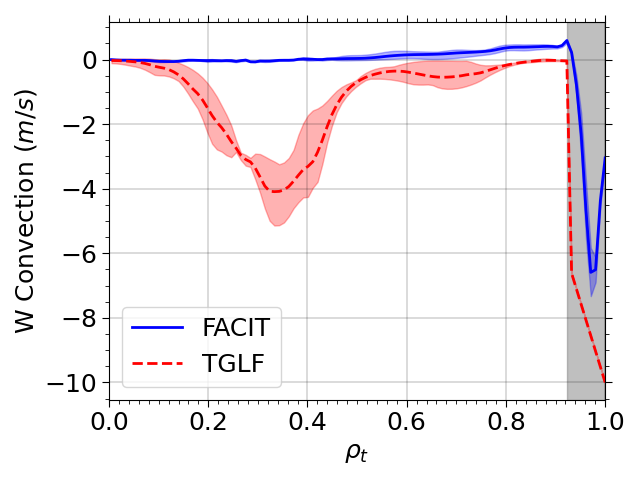}
    \includegraphics[width=0.45\textwidth]{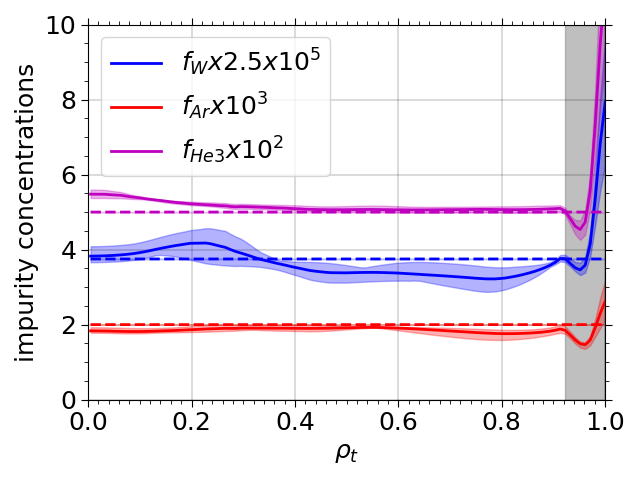}
    \includegraphics[width=0.45\textwidth]{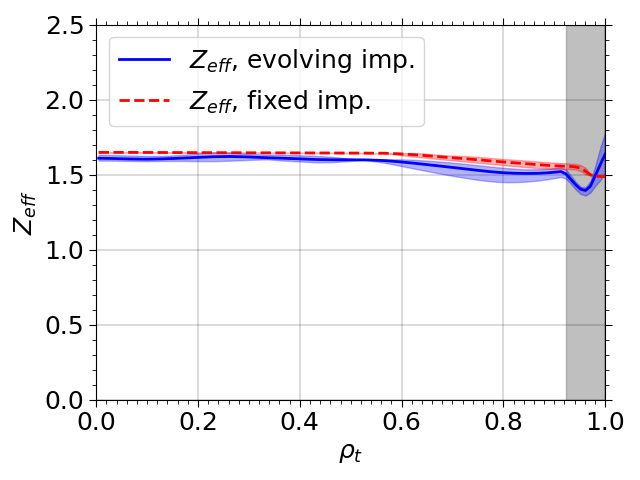}
    \caption{8T H-mode profiles. Top left: turbulent (red) and neoclassical (blue) W diffusivity; top right: turbulent (red) and neoclassical (blue) W convection; bottom left: impurity concentrations; bottom right: $Z_{eff}$. In the bottom plots, the dashed lines indicate the simulations with fixed concentrations. The colored shades indicate the maximum and minimum values computed across the scans. The gray region indicates from top of pedestal to the separatrix.}
    \label{fig:H8_imp_profiles}
\end{figure}

The low magnetic field H-mode (H8) is a scenario considered for SPARC early operation. Its main parameters are listed in table \ref{tab:8T}.
\begin{table}[h]
  \centering
  \caption{List of the main engineering parameters of the SPARC 8T H-mode discharge. Their definition is in \cite{creely_overview_2020}.}
  \begin{tabular}{cc}
    \toprule
    \rowcolor{lightblue} 
    \textbf{Parameter} & \textbf{Value} \\
    \midrule
          $B_t$  &  8 T  \\
          $I_p$  &  5.7 MA  \\
          $k_{sep}$  &  2  \\
          $\delta_{sep}$  &  0.54  \\
          $q^*_{Uckan}$  &  3  \\
          $Z_{eff}$  &  1.5  \\
    \bottomrule
  \end{tabular}
  \label{tab:8T}
\end{table}
This H-mode operates at a lower $B_t=8T$ constrained by the ICRH heating technique at 120 MHz based on near-axis power absorption of a H minority species. To maintain the same $q^*$, the plasma current has been reduced. Such variations of global parameters are expected to reflect lower performance than the PRD, making this scenario more suitable for the early operation of SPARC.
For this pulse a density scan has been performed, following the same methodology used for the PRD, and assuming $P_{ICRH}=25MW$. The results of the scan are shown in figure \ref{fig:H8_scan}.
This figure shows an almost perfect match of fusion gain, validating the simplified approach with fixed impurity concentrations. $f_{LH}$ shows instead an offset between the two simulation approaches. In particular, the cases with impurity transport are more optimistic about H-mode accessibility. The reason behind the mismatch is the difference in the predicted ion heat flux, which affects $f_{LH}$, using the Schmidtmayr scaling. The different $Q_i$ is caused by the different radiation calculated in the simulations.
\begin{figure}[h]
    \centering
    \includegraphics[width=0.45\textwidth]{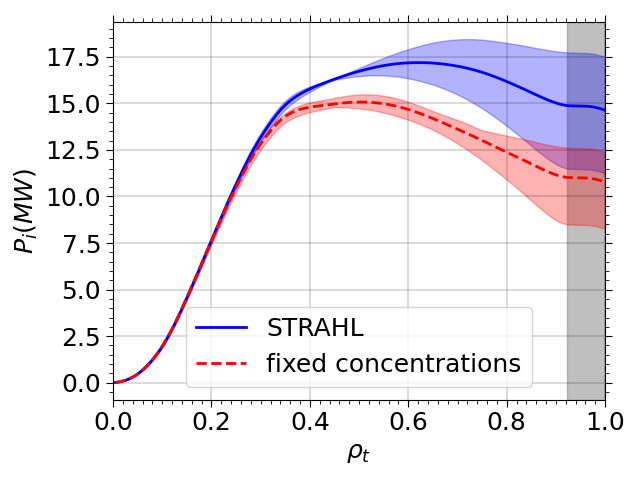}
    \includegraphics[width=0.45\textwidth]{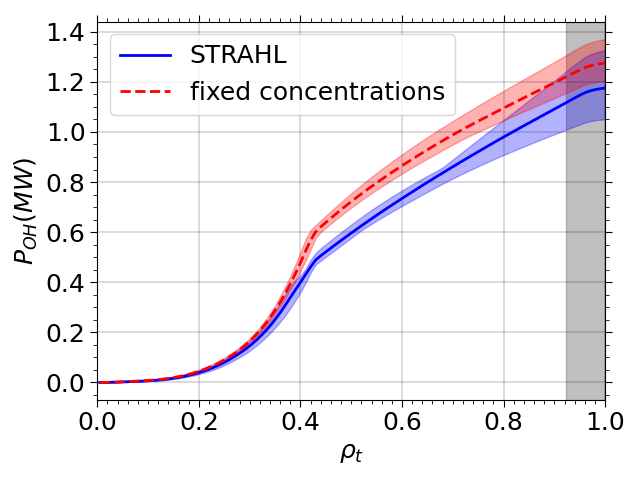}
    \includegraphics[width=0.45\textwidth]{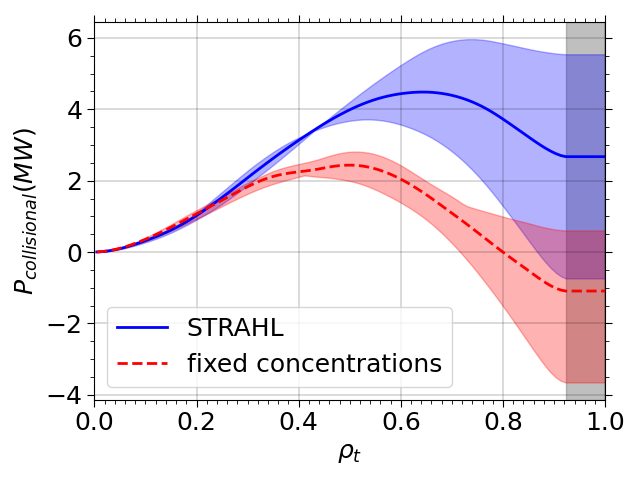}
    \includegraphics[width=0.45\textwidth]{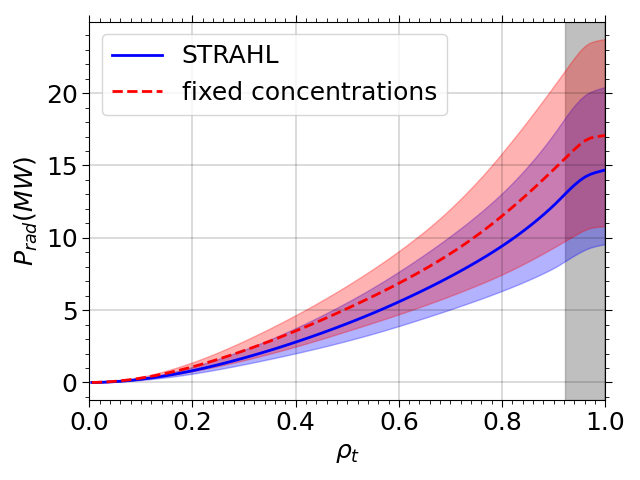}
    \includegraphics[width=0.45\textwidth]{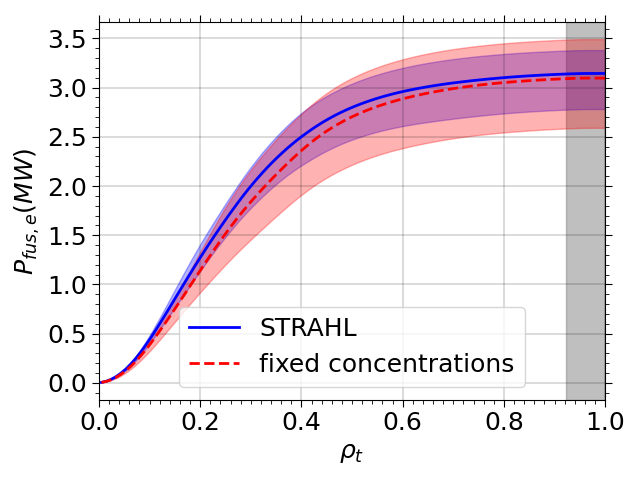}
    \includegraphics[width=0.45\textwidth]{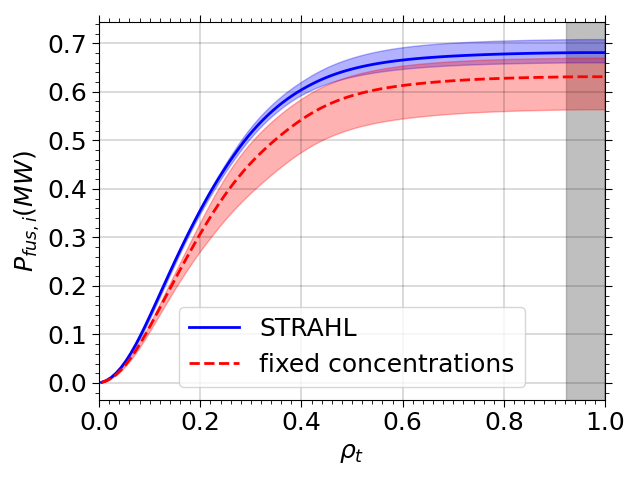}
    \caption{Integrated radial profiles of the power fluxes for a density scan of the SPARC H8 scenario, using STRAHL (blue) and fixed impurity concentrations (red). Top left: total ion power; top right: ohmic power; center left: collisional exchange; center right: radiative power; bottom left: fusion power to electrons; bottom right: fusion power to ions. The solid/dashed lines indicate the nominal simulations, while the shaded area show the maximum and minimum values of the profiles across the density scan. The gray region indicates from top of pedestal to the separatrix.}
    \label{fig:H8_powerfluxes}
\end{figure}
\begin{figure}[h]
    \centering
    \includegraphics[width=0.45\textwidth]{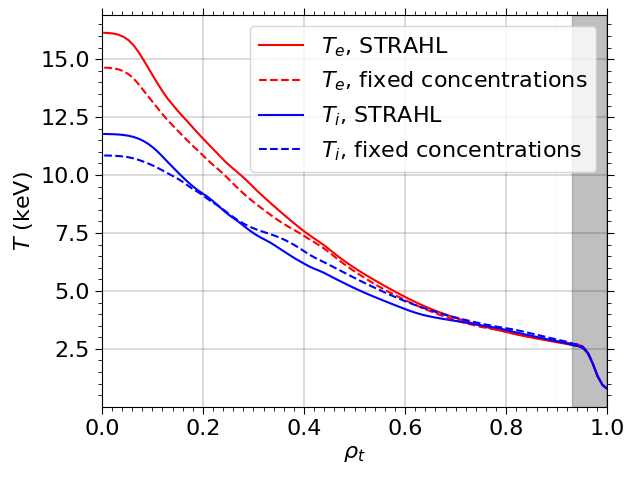}
    \caption{Temperature profiles at the lowest density of the SPARC H8 scenario, using STRAHL (solid) and fixed impurity concentrations (dashed). Red (blue) indicates the electron (ion) temperature. The gray region indicates from top of pedestal to the separatrix.}
    \label{fig:H8_Tprofiles}
\end{figure}
\begin{figure}[h]
    \centering
    \includegraphics[width=0.45\textwidth]{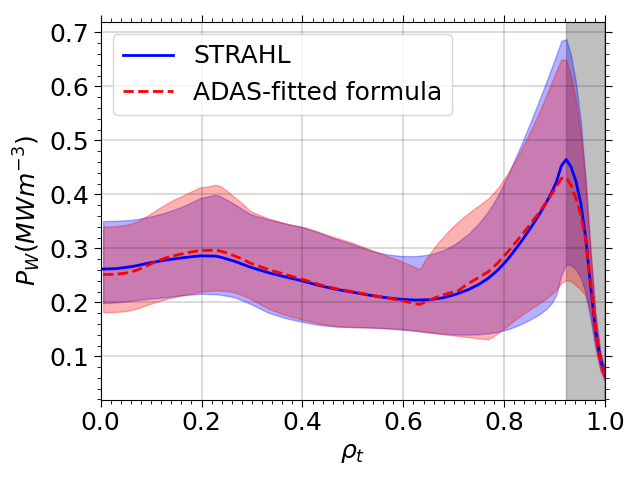}
    \includegraphics[width=0.45\textwidth]{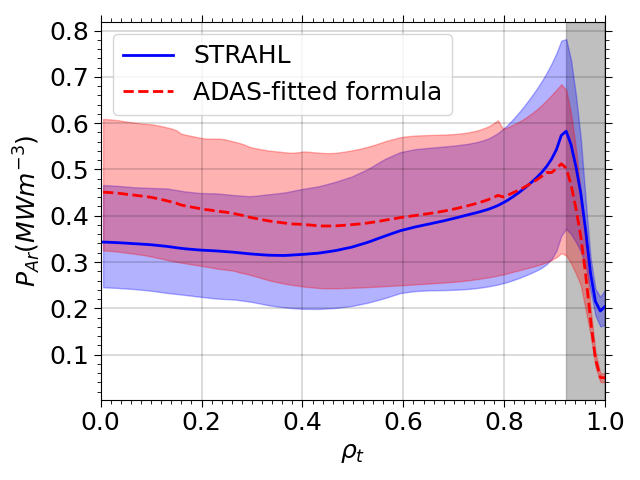}
    \includegraphics[width=0.45\textwidth]{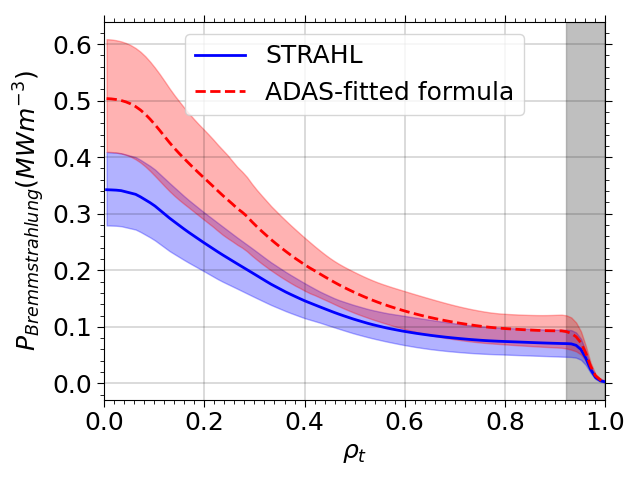}
    \caption{Radiative power density profiles for a density scan of the SPARC H8 scenario, using STRAHL (blue) and fixed impurity concentrations (red). Top left: W radiation; top right: Ar radiation; bottom: Bremsstrahlung radiation. The solid/dashed lines indicate the nominal simulations, while the shaded area show the maximum and minimum values of the profiles across the density scan. The gray region indicates from top of pedestal to the separatrix.}
    \label{fig:H8_radiations}
\end{figure}
In fact, in figure \ref{fig:H8_powerfluxes}, one can see that most of the power fluxes are similar using STRAHL or fixing the impurity concentrations, but a deviation of the radiative power causes a lower electron temperature which is reflected in a smaller collisional exchange from electrons to ions. The temperature profiles for the case at lowest density are shown in figure \ref{fig:H8_Tprofiles}. The lower collisional exchange causes lower $Q_i$ in the simulations with fixed concentrations. In order to verify the origin of the mismatch in the radiative power, $P_{rad}$ from Bremsstrahlung, W and Ar has been plotted in figure \ref{fig:H8_radiations}. The figure shows a great match for W and a small deviation for Ar, while the biggest discrepancy is found for the Bremsstrahlung radiation. In the simulations without STRAHL, $P_{rad,W}$ and $P_{rad,Ar}$ are calculated respectively as in \cite{putterich_calculation_2010} an \cite{post_steady-state_1977}, while $P_{Bremsstrahlung}=5.06\times10^{-5}Z_{eff}n_e^2\sqrt{T_e}$, as in \cite{uckan_iter_1991}. The maximum $P_{Bremsstrahlung}$ deviation is 33\%, which is reflected also in $f_{LH,Schmidtmayr}$ in figure \ref{fig:H8_scan} , showing that the simplified formula used in the simulations with fixed concentration can overestimate the radiative power. For comparison, the H-mode robustness calculated with the Martin scaling \cite{martin_power_2008} has been analyzed as well. Table \ref{tab:8T_fLH} compares $f_{LH,STRAHL}/f_{LH,fixed}$ using Martin and Schmidtmayr scalings, where $f_{LH,STRAHL}$ is calculated in the simulations with STRAHL, and $f_{LH,fixed}$ in the cases with fixed impurity concentrations.
\begin{table}[h]
  \centering
  \caption{$f_{LH,STRAHL}/f_{LH,fixed}$ for different Greenwald fractions, using Martin or Schmidtmayr scaling.}
  \begin{tabular}{c c c}
    \toprule
    \rowcolor{orange}
     \textbf{$f_G$} & \textbf{$(f_{LH,STRAHL}/f_{LH,fixed})_{Martin}$} & \textbf{$(f_{LH,STRAHL}/f_{LH,fixed})_{Schmidtmayr}$} \\
     0.39 & 1.06 & 1.3 \\
     0.48 & 1.2 & 1.27 \\
     0.56 & 1.5 & 1.33 \\
    \bottomrule
  \end{tabular}
  \label{tab:8T_fLH}
\end{table}
The table shows higher $f_{LH}$ values in STRAHL simulations with both scalings. However, while $(f_{LH,STRAHL}/f_{LH,fixed})_{Schmidtmayr}$ is constant across the scan, $(f_{LH,STRAHL}/f_{LH,fixed})_{Martin}$
increases with density, showing high discrepancy at maximum $f_G$. The reason behind this behavior is the crucial role that radiation plays in the electron heat flux, which is included in the $f_{LH,Martin}$ computation. In fact, when using the Martin scaling, $Q_e$ is directly affected by $P_{rad}$, while $Q_i$ and $f_{LH,Schmidtmayr}$ are influenced only indirectly through modifications of the electron temperature and collisional exchange. The discrepancy in $f_{LH}$ was not observed for the PRD, because for such scenario the fusion power is significantly higher, making the radiation and collisional exchange contributions small at the W concentrations predicted in the simulations.
The neoclassical pinch and turbulent diffusivity of the simulations are shown in figure \ref{fig:H8_imp_profiles}, together with the impurity concentrations.
As found for the PRD, the W and Ar turbulent transport prevails on the neoclassical one and the impurity concentrations are mostly flat, matching the simplified simulations.

\subsection{Reduced Plasma Current, $B_t=12~T$}
\begin{figure}
    \centering
    \includegraphics[width=0.45\linewidth]{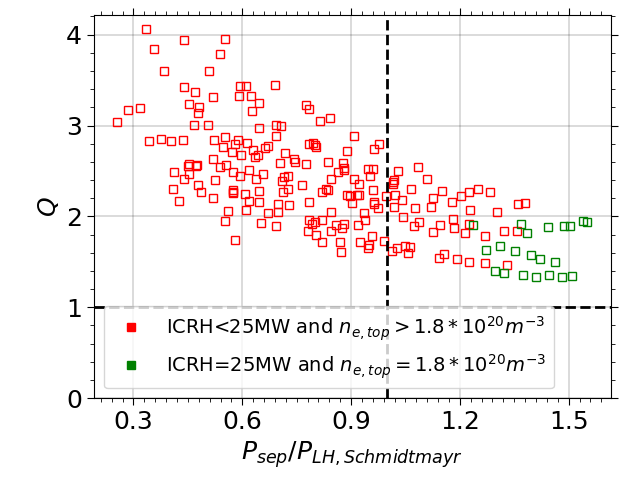}
    \includegraphics[width=0.45\linewidth]{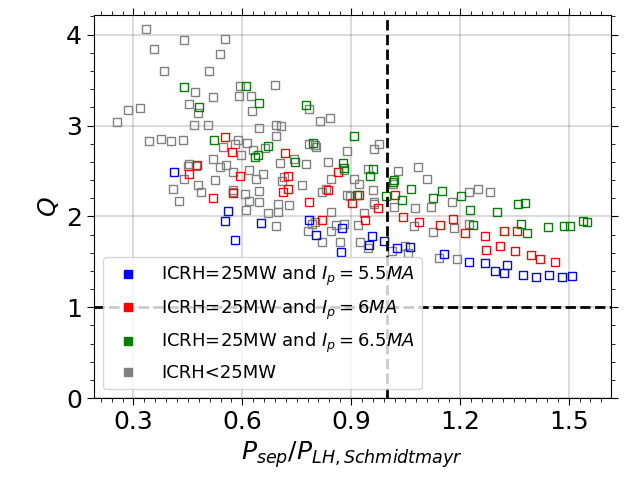}
    \caption{Left: $Q$ vs $f_{LH}$ for the H12 scenario. The green squares are simulations with maximum available $P_{ICRH}$ and lowest top of pedestal density. Right: $Q$ vs $f_{LH}$ for maximum ICRH input power and different $I_p$ values. Gray squares are simulations with $P_{ICRH}<25MW$ , at any plasma current.}
    \label{fig:H12_Q_vs_fLH}
\end{figure}

In an attempt to identify non-L-mode scenarios that reach breakeven conditions, we have investigated an additional H-mode operated with 12 T and reduced plasma current. Such pulse has been found through a scoping of the operational space, exploring different values of the engineering and global parameters. An initial broad scope has been performed with Plasma OPerational CONtour (POPCON) \cite{houlberg_contour_1982}, which couples 0D models (e.g. scaling laws), functional forms to describe kinetic profiles, and physical/technological constraints (e.g. $q^*$ or $q_{95}$) to find scenarios which meet certain requirements, as a specific fusion gain or power. This has been done with the additional constraint of $f_{LH}>1$ (i.e. accessing H-mode according to both Martin and Schmidtmayr scalings ). The results produced the global parameters listed in table \ref{tab:H12}.
\begin{table}[h]
  \centering
  \caption{List of the main engineering parameters of the reduced current H-mode discharge. Their definition is in \cite{creely_overview_2020}.}
  \begin{tabular}{cc}
    \toprule
    \rowcolor{lightblue} 
    \textbf{Parameter} & \textbf{Value} \\
    \midrule
          $B_t$  &  12.2 T  \\
          $I_p$  &  6.5 MA  \\
          $k_{sep}$  &  2  \\
          $\delta_{sep}$  &  0.48  \\
          $q^*_{Uckan}$  &  3.6  \\
          $Z_{eff}$  &  3.0  \\
    \bottomrule
  \end{tabular}
  \label{tab:H12}
\end{table}
It is worth to notice the higher $Z_{eff}$ compared to the previously analyzed H-modes. Such estimate is reflected by a heavy SOL Ar seeding aimed at protecting the divertor and PFCs at the high $q_{95}$ conditions reached here.
After the initial POPCON exploration, an additional scoping has been performed with ASTRA, using fixed impurity concentrations, and performing a 4D scan of density at top of pedestal, ICRH input power, W concentration and plasma current. The ranges of variation of such parameters are listed in table \ref{tab:H12_ranges}.
\begin{table}[h]
  \centering
  \caption{Ranges of variation of engineering and global parameters for the reduced current H-mode.}
  \begin{tabular}{cc}
    \toprule
    \rowcolor{brown} 
    \textbf{Parameter} & \textbf{Value} \\
    \midrule
          $I_p$  &  [5.5 - 6.5] $MA$ \\
          $P_{ICRH}$  &  [15 - 25] $MW$ \\
          $n_{top}$  &  [1.8 - 3.3] $10^{20}m^{-3}$ \\
          $f_W$  &  [1.5 - 13.5] $10^{-5}$ \\
    \bottomrule
  \end{tabular}
  \label{tab:H12_ranges}
\end{table}
With these permutations a total amount of 657 simulations was obtained, largely spanning the operational space. As found in \cite{muraca_integrated_2025} for the PRD and the 8T H-modes, the conditions which maximize core fusion gain and H-mode access probability are low density and high ICRH input power. The effect of such conditions on the exhaust and PFCs protection has not been investigated here. The results of this exploration are shown in figure \ref{fig:H12_Q_vs_fLH}, where the suitable operational window for breakeven campaign is where $Q>1$ and $f_{LH}>1$. In the left plot, the green symbols indicate simulations at the lowest density and highest ICRH input power. In the right plot, an additional result is shown: the breakeven H-mode sustained condition can be reached at different levels of plasma current. This ensures a certain flexibility in the choice of $q_{95}$, allowing for multiple strategies to handle exhaust power.
\begin{figure}
    \centering
    \includegraphics[width=0.45\linewidth]{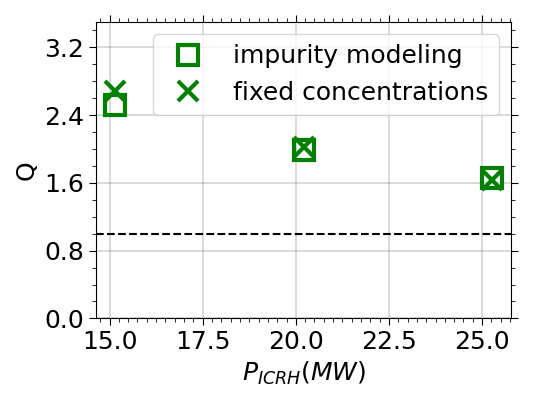}
    \includegraphics[width=0.45\linewidth]{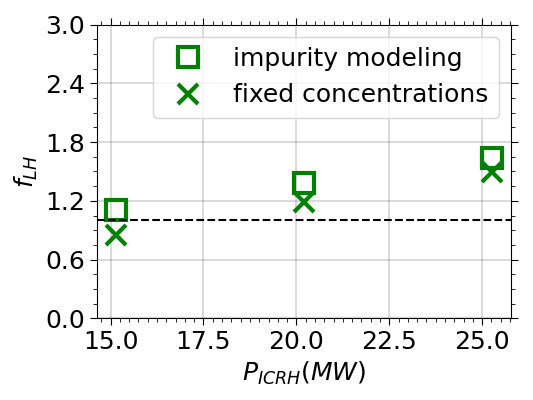}
    \includegraphics[width=0.45\linewidth]{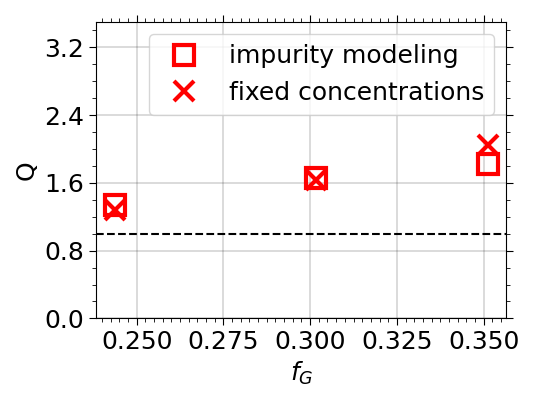}
    \includegraphics[width=0.45\linewidth]{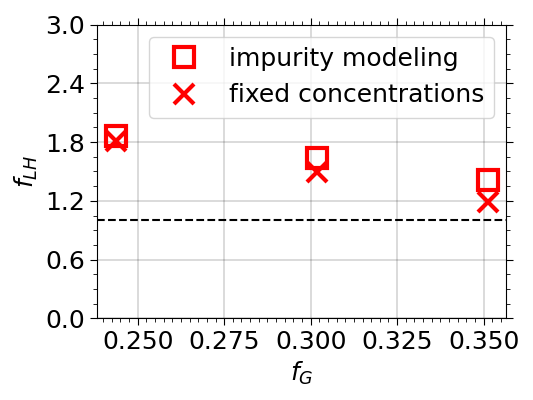}
    \caption{On the left and right are shown respectively fusion gain and $f_{LH}$ values for an ICRH power (top plots, green) and density (bottom plots, red) scan of the H12 scenario. The squares indicate simulations with impurity transport, while the crosses are simulated keeping fixed radial impurity concentrations.}
    \label{fig:H12_scan}
\end{figure}
\begin{figure}
    \centering
    \includegraphics[width=0.45\linewidth]{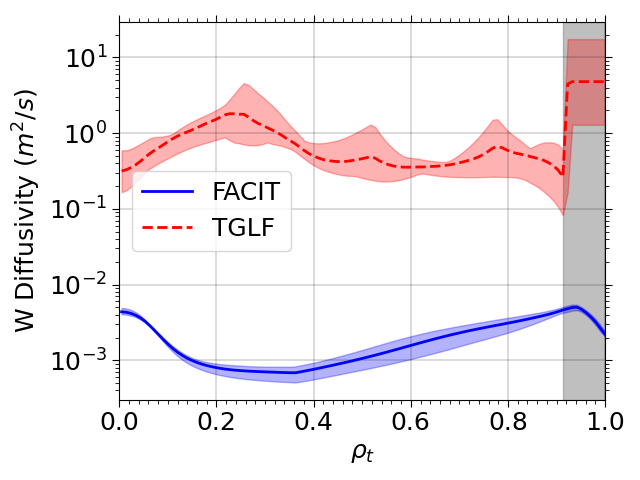}
    \includegraphics[width=0.45\linewidth]{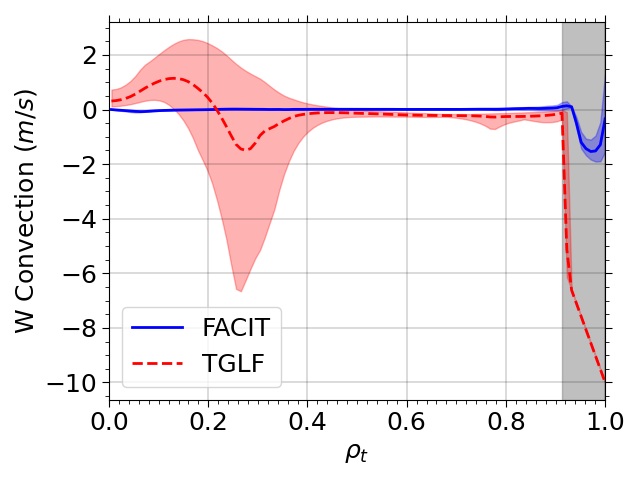}
    \includegraphics[width=0.45\linewidth]{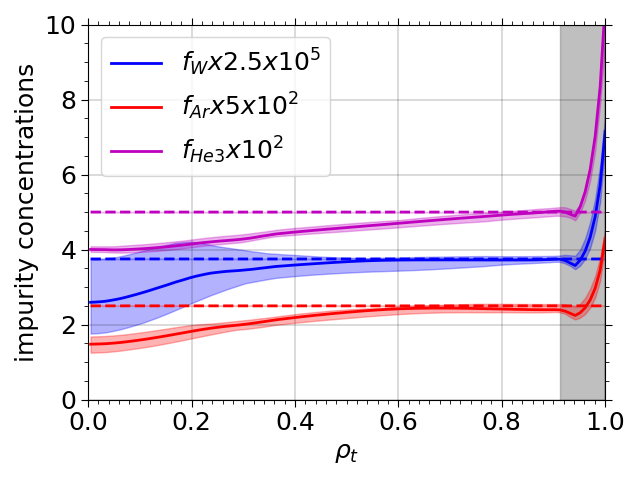}
    \includegraphics[width=0.45\linewidth]{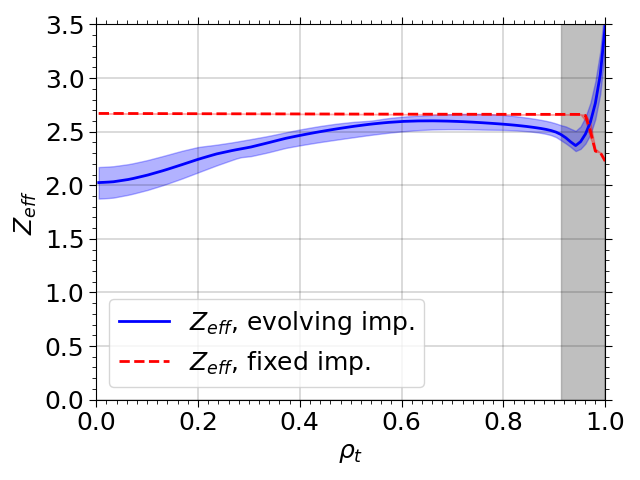}
    \caption{Low current H-mode profiles. Top left: turbulent (red) and neoclassical (blue) W diffusivity; top right: turbulent (red) and neoclassical (blue) W convection; bottom left: impurity concentrations; bottom right: $Z_{eff}$. In the bottom plots, the dashed lines indicate the simulations with fixed concentrations. The colored shades indicate the maximum and minimum values computed across the scans. The gray region indicates from top of pedestal to the separatrix.}
    \label{fig:H12_imp_profiles}
\end{figure}
The scoping of the operational space performed with ASTRA has identified ideal values of density and ICRH power at a certain prescribed W concentration. The results should then be compared with higher-fidelity simulations which include impurity transport, as the ones previously performed for PRD and H8. The results of such comparison are shown in figure \ref{fig:H12_scan}, showing a top of pedestal and ICRH power scan. For this scenario a very good benchmark has been obtained, with a small deviation of $f_{LH}$, similarly to H8. The turbulent diffusivity and neoclassical pinch are plotted in figure \ref{fig:H12_imp_profiles} for Ar and W, together with impurity concentrations and $Z_{eff}$. The figure shows that the W anomalous transport is considerably higher than the neoclassical pinch, and as a consequence the radial W concentration is roughly constant. However, He3 and Ar show slightly hollowed concentration profiles, showing that the assumption of fixed concentrations is conservative for radiation and $f_{LH}$ calculations. The hollowed Ar profile is reflected also by $Z_{eff}$, as shown in figure \ref{fig:H12_imp_profiles}.

\section{Sensitivity of impurity peaking to Ar and W pedestal concentrations}
\begin{figure}[h]
    \centering
    \includegraphics[width=0.49\linewidth]{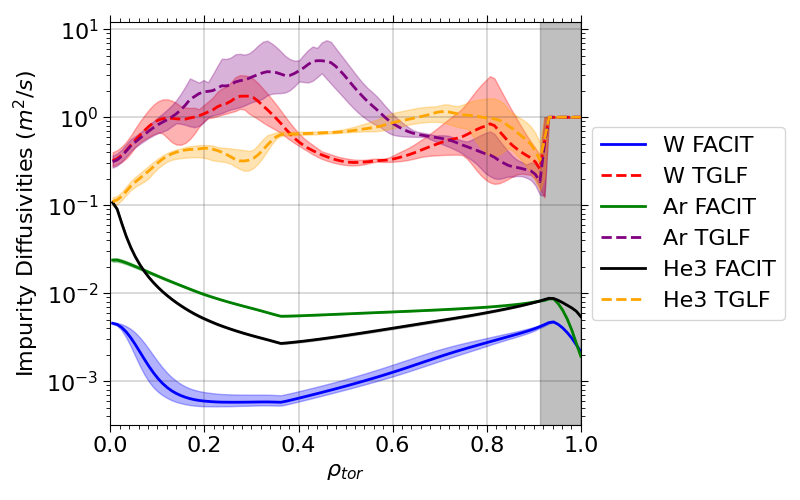}
    \includegraphics[width=0.49\linewidth]{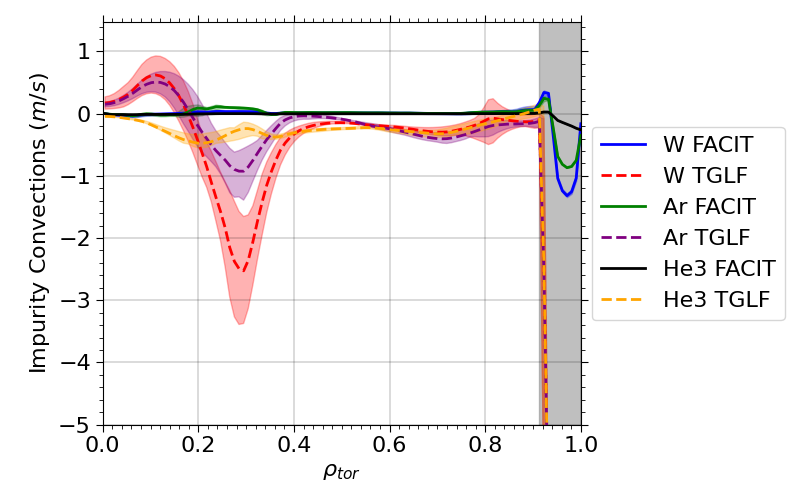}
    \caption{Scan of W pedestal concentration for the H12 scenario. Profiles of turbulent (dashed) and neoclassical (solid) transport coefficients. On the left (right) the diffusivities (convections) are shown for all the impurities. The colored shades indicate the maximum and minimum values computed for the transport coefficients across the scan. The gray area is from top of pedestal to the separatrix.}
    \label{fig:Wscan_profiles}
\end{figure}

\begin{figure}[h]
    \centering
    \includegraphics[width=0.45\linewidth]{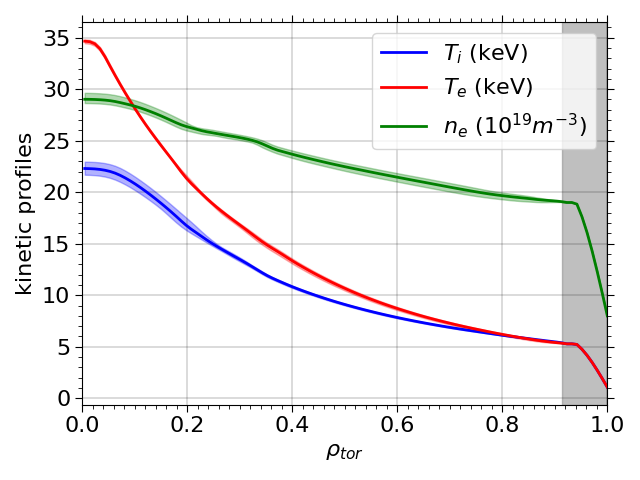}
    \includegraphics[width=0.45\linewidth]{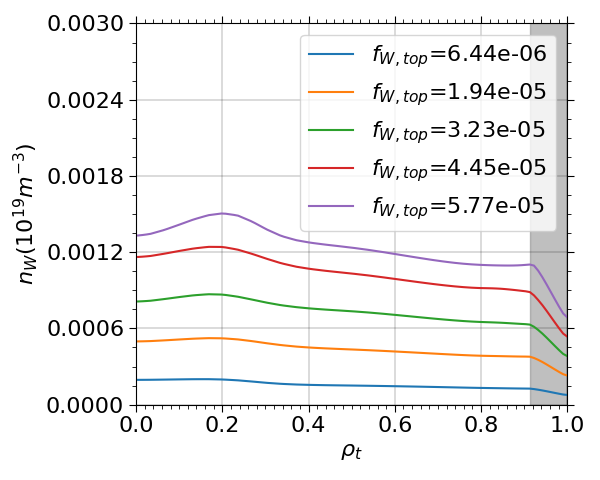}
    \includegraphics[width=0.45\linewidth]{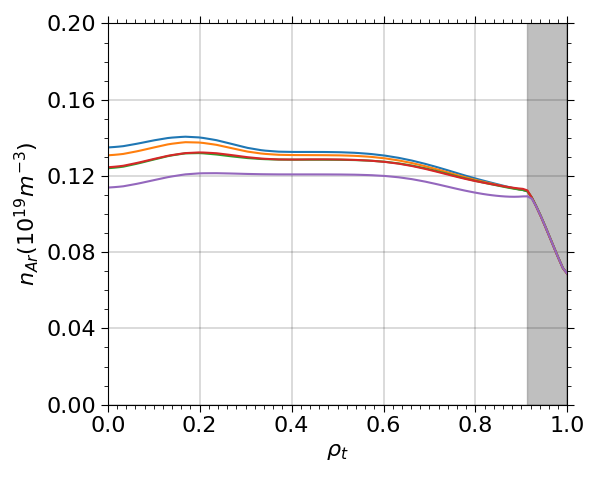}
    \includegraphics[width=0.45\linewidth]{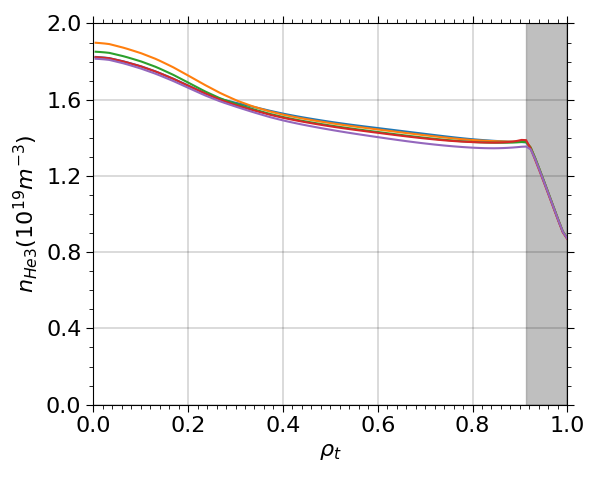}
    \caption{Profiles for a scan of W pedestal concentration in the H12 scenario. Top left: kinetic profiles; top right: W density; bottom left: Ar density; bottom right: He3 density. The colored shades (top left) indicate the maximum and minimum profiles computed across the scan. The gray area is from top of pedestal to the separatrix.}
    \label{fig:Wscan_kinprofiles}
\end{figure}

\begin{figure}[h]
    \centering
    \includegraphics[width=0.49\linewidth]{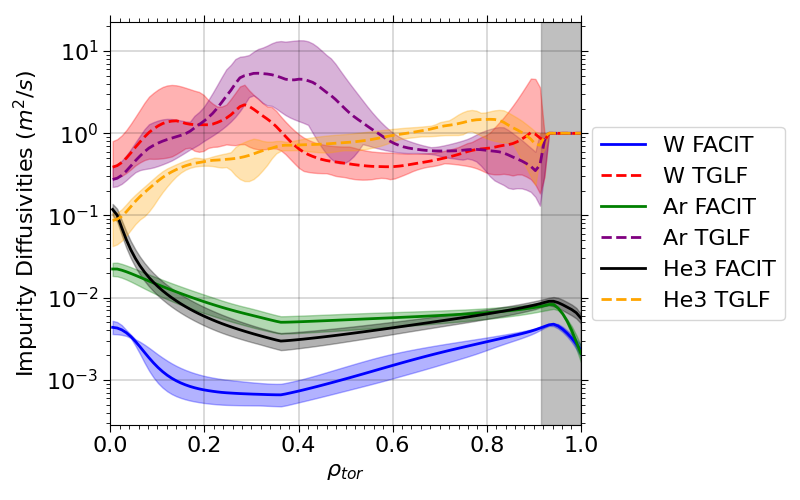}
    \includegraphics[width=0.49\linewidth]{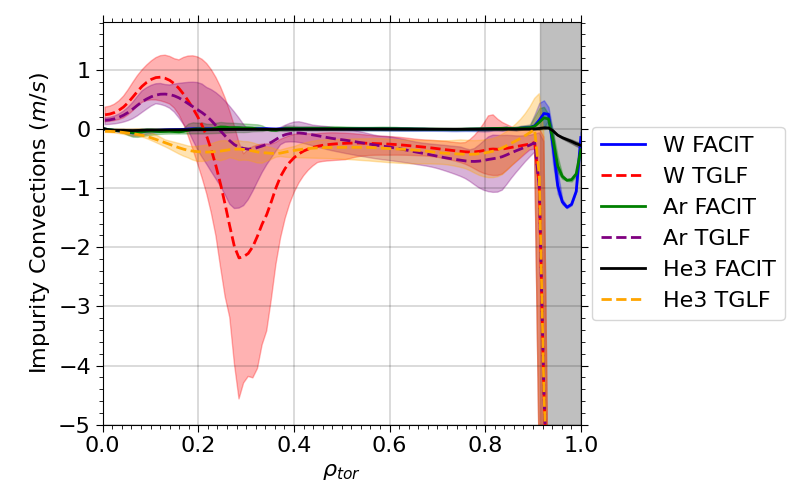}
    \caption{Scan of Ar pedestal concentration in the H12 scenario. Profiles of turbulent (dashed) and neoclassical (solid) transport coefficients. On the left (right) the diffusivities (convections) are shown for all the impurities. The colored shades indicate the maximum and minimum values computed for the transport coefficients across the scan. The gray area is from top of pedestal to the separatrix.}
    \label{fig:Arscan_profiles}
\end{figure}

\begin{figure}[h]
    \centering
    \includegraphics[width=0.49\linewidth]{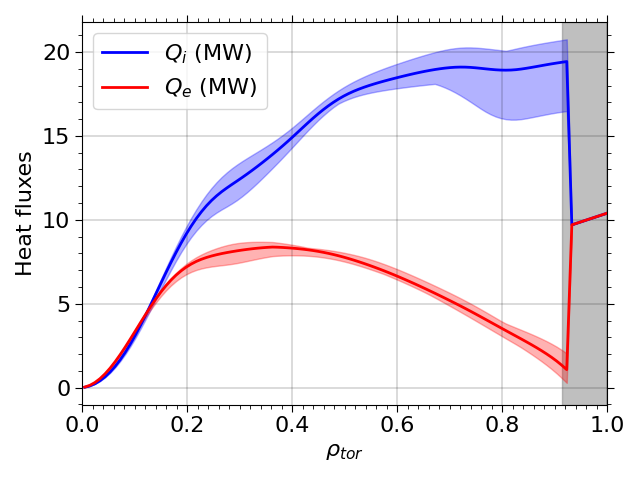}
    \includegraphics[width=0.49\linewidth]{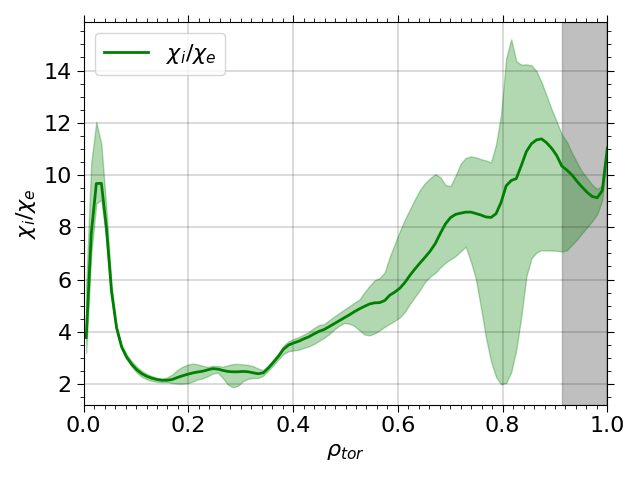}
    \caption{On the left the ion (blue) and electron (red) heat fluxes are shown, while on the right $\chi_i/\chi_e$ is pictured. The colored shades indicate the maximum and minimum values computed across the W concentration scan. The gray area is from top of pedestal to the separatrix.}
    \label{fig:Wscan_heatfluxes}
\end{figure}

\begin{figure}[h]
    \centering
    \includegraphics[width=0.45\linewidth]{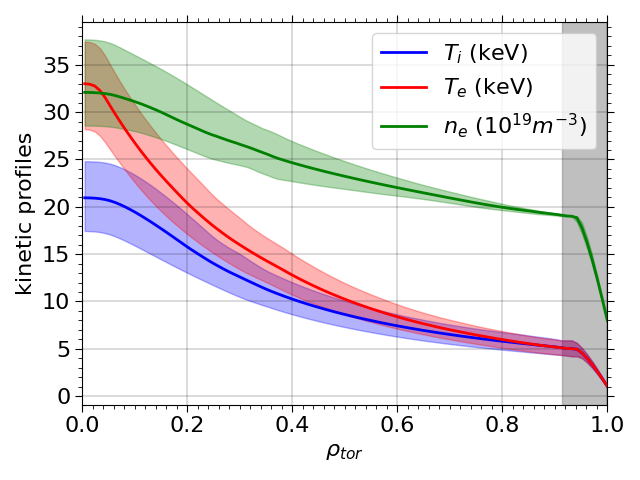}
    \includegraphics[width=0.45\linewidth]{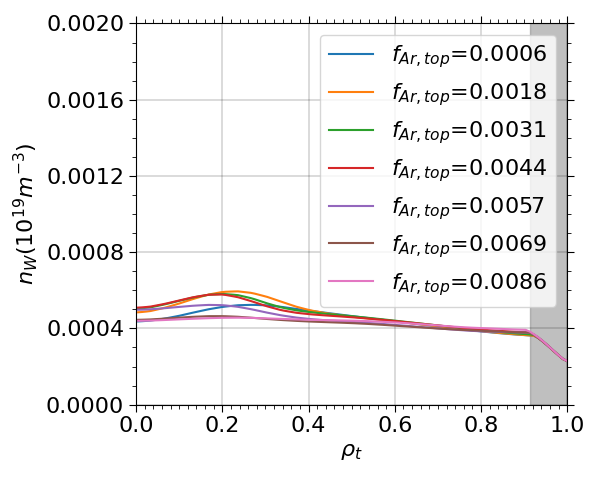}
    \includegraphics[width=0.45\linewidth]{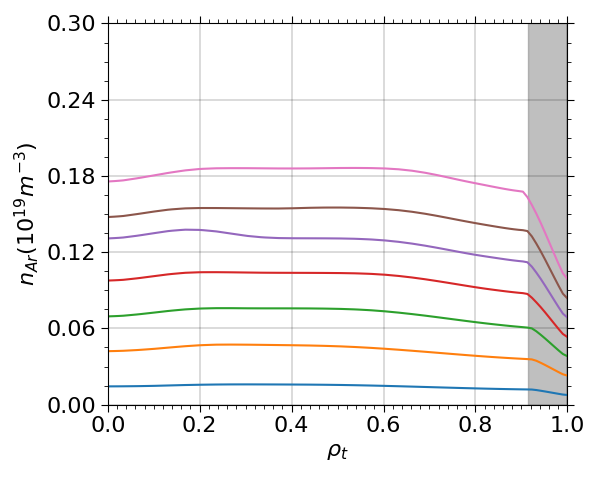}
    \includegraphics[width=0.45\linewidth]{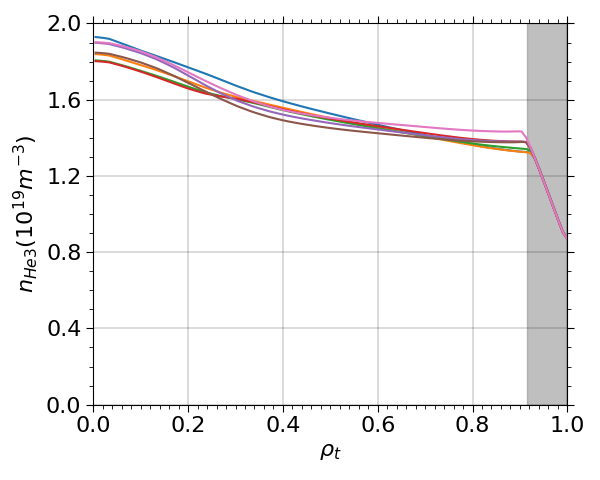}
    \caption{Profiles for a scan of Ar pedestal concentration in the H12 scenario. Top left: kinetic profiles; top right: W density; bottom left: Ar density; bottom right: He3 density. The colored shades (top left) indicate the maximum and minimum profiles across the scan. The gray area is from top of pedestal to the separatrix.}
    \label{fig:Arscan_kinprofiles}
\end{figure}

\begin{figure}[h]
    \centering
    \includegraphics[width=0.45\linewidth]{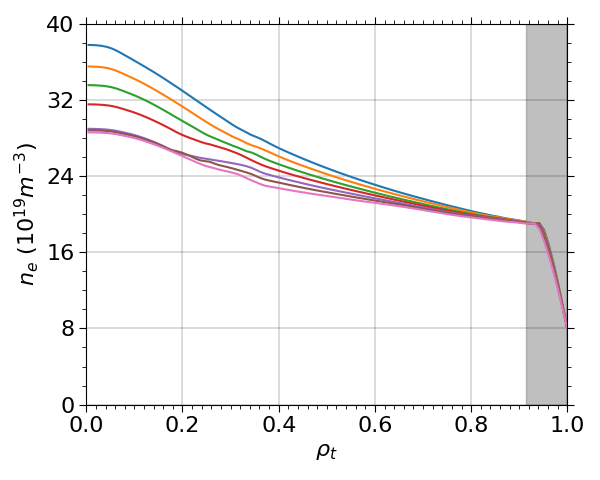}
    \includegraphics[width=0.45\linewidth]{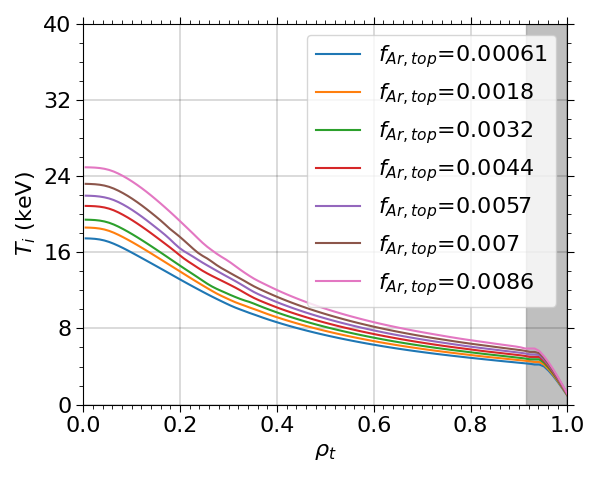}
    \includegraphics[width=0.45\linewidth]{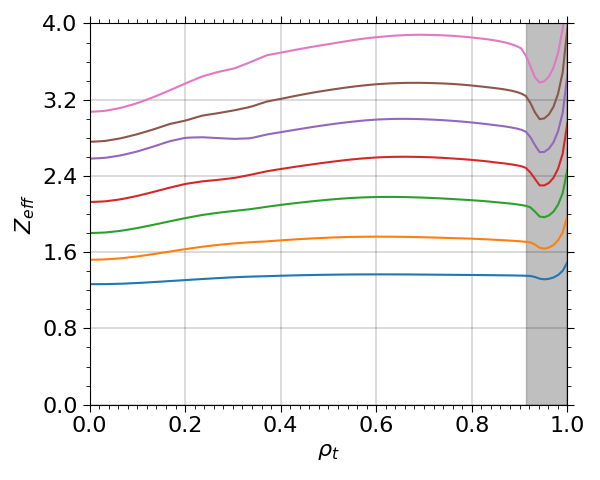}
    \includegraphics[width=0.45\linewidth]{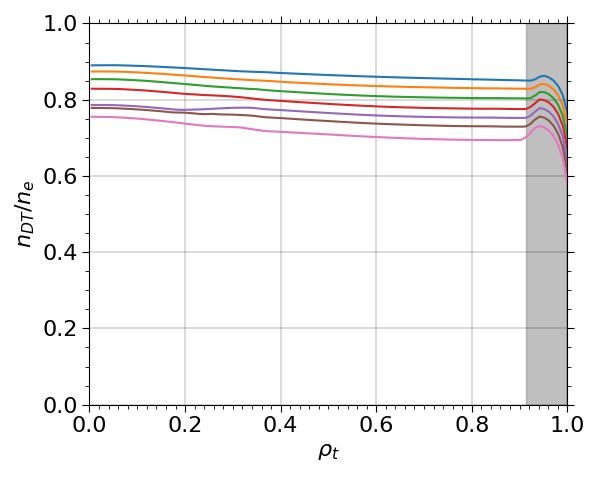}
    \includegraphics[width=0.45\linewidth]{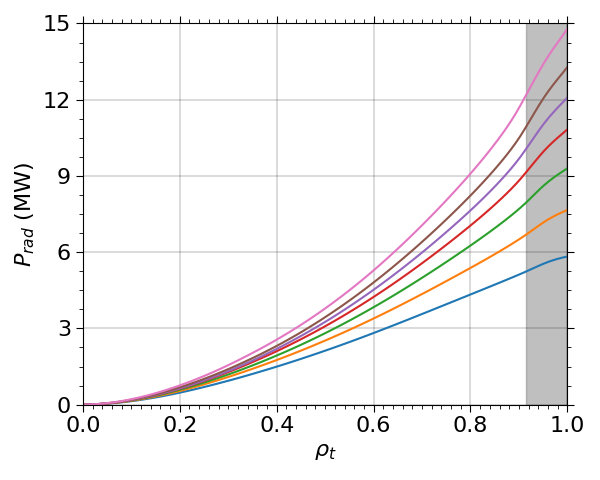}
    \includegraphics[width=0.45\linewidth]{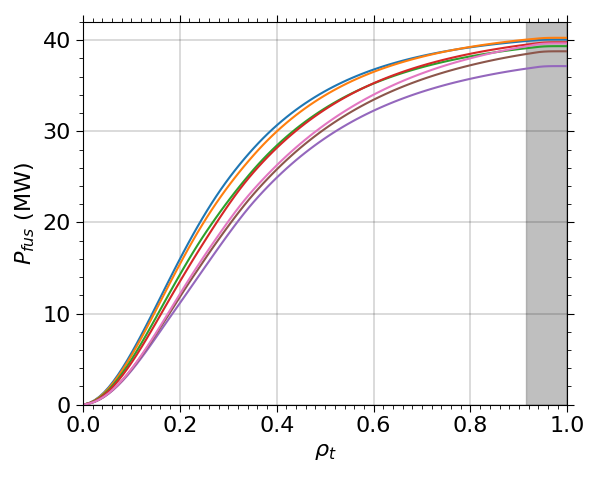}
    \caption{Profiles for a scan of Ar pedestal concentration in the H12 scenario. Top left: electron density; top right: ion temperature; center left: effective charge; center right: DT concentration; bottom left: cumulative radiative power; bottom right: total fusion power. The gray area is from top of pedestal to the separatrix.}
    \label{fig:Arscan_otherprofiles}
\end{figure}

The simulations shown in this work assumed a set of boundary conditions for the impurity concentrations at the top of pedestal. This choice is justified by the difficulty of modeling turbulent transport in the edge of a H-mode. However, the concentrations at the pedestal position are uncertain. Therefore, for the H12 scenario, scans in Ar and W concentrations have been performed to study the impact on the impurity peaking, H-mode robustness and fusion performance. These concentrations have been obtained scanning the impurity sources from the wall in STRAHL and keeping a fixed turbulent impurity transport in the edge. The ranges of variation of the scans are listed in table \ref{tab:H12_scans}.
\begin{table}[h]
  \centering
  \caption{Ranges of variation of W and Ar top of pedestal concentrations for the H12 scenario.}
  \begin{tabular}{>{\columncolor{brown}}c c} 
    \toprule
    \rowcolor{brown} 
     & \textbf{Concentration} \\
    \midrule
    \textbf{W scan} & $f_{W,top}$ = [0.6 -- 6] $\times 10^{-5}$\\
    \textbf{Ar scan} & $f_{Ar,top}$ = [0.5 -- 8] $\times 10^{-3}$\\
    \bottomrule
  \end{tabular}
  \label{tab:H12_scans}
\end{table}
Increasing the W concentration, an effect on the H-mode sustainability has been found, as expected, due to the higher W radiation. However, a minimal impact has been observed on the fusion performance and impurity peaking. The impurity transport coefficients are plotted in figure \ref{fig:Wscan_profiles} for all species.
The kinetic profiles are plotted in figure \ref{fig:Wscan_kinprofiles}, together with the impurity densities.
One can notice that there is almost no effect on the kinetic profiles. This is due to the stiffness of the kinetic profiles \cite{imbeaux_modelling_2001,ryter_electron_2003} which characterizes the high power H-mode scenario analyzed here. The left plot of figure \ref{fig:Wscan_heatfluxes} shows a small deviation of the electron and ion heat fluxes, due to the different radiation contribution. On the right plot of the same figure, $\chi_i/\chi_e$ is pictured, showing average values consistent with typically ITG-dominated regimes \cite{kotschenreuther_gyrokinetic_2019}. The Ar and He3 profiles show minimal variations, decreasing their average densities at higher W concentrations. Such variations are probably associated with the slightly higher $Z_{eff}$ reached at higher $f_W$ values. However, since the $f_W$ scan (larger than an order of magnitude) has produced a local maximum variation of ~8\% in the profiles, further investigation has not been considered necessary. \\
The scan in Ar concentration provided interesting results. The impurity transport coefficients are shown in figure \ref{fig:Arscan_profiles}, while the kinetic and impurity density profiles are shown in figure \ref{fig:Arscan_kinprofiles}.
Figure \ref{fig:Wscan_profiles} and \ref{fig:Arscan_profiles} confirm that the anomalous impurity transport prevails on the neoclassical one in the core, as was previously found for all the performed simulations, including PRD and H8 scenarios.
In figure \ref{fig:Arscan_kinprofiles}, the kinetic profiles show a non-negligible variation across the Ar scan. In fact, a variation of Ar content affects multiple physics processes at the same time. In particular, a higher $f_{Ar}$:
\begin{itemize}
    \item leads to a higher $Z_{eff}$. This has a double effect: first, it impacts the ITG/TEM core instability balance. In fact, a higher $Z_{eff}$ increases the ITG linear stability threshold \cite{jenko_critical_2001}, making the plasma more TEM-dominated, condition which typically shows a lower density peaking \cite{fable_role_2010}; second, it increases the top of pedestal pressure (i.e. higher temperatures at fixed input density), since for the simulated electron density the pedestal stability is peeling-constrained;
    \item increases the main ion (DT) dilution, affecting turbulence \cite{ennever_effects_2015, kim_full-f_2017, rodriguez-fernandez_core_2024} and leading to a smaller amount of fuel for fusion reactions;
    \item increases the radiative power, reducing the H-mode robustness.
\end{itemize}
All the effects discussed above can be observed in figure \ref{fig:Arscan_otherprofiles}.
While the higher radiation has a clear impact on H-mode sustainment, the overall effect on fusion performance is less trivial. In fact, the higher pedestal facilitates the production of higher fusion power, but the lower density peaking reduces $P_{fus}$ and the lower DT concentration reduces the available fuel for fusion reactions. As a result of all the competing terms, the fusion power is roughly constant, as can be seen at the bottom right plot of figure \ref{fig:Arscan_otherprofiles}. It is worth to mention that a similar competition between positive and negative effects has not been found across the W scan, since its low concentration makes it a trace species, not affecting considerably $Z_{eff}$ and the DT dilution.
A few global parameters to summarize the results of the W and Ar concentration scans are respectively listed in table \ref{tab:H12_Wscan_summary} and \ref{tab:H12_Arscan_summary}.
\begin{table}[h]
  \centering
  \caption{Global parameters for the scan in W pedestal concentration of the H12 scenario.}
  \begin{tabular}{>{\columncolor{green}}c c c c c} 
    \toprule
    \rowcolor{green} 
     \textbf{$f_{W,top}$ $(10^{-5})$} & \textbf{$<T_i>$ $(keV)$} & \textbf{$P_{rad}$ $(MW)$} & \textbf{$Q$} & \textbf{$f_{LH,Schmidt}$}\\
    \midrule
     0.6 & 8 & 10.7 & 1.43 & 1.58 \\
     2 & 8.05 & 12 & 1.43 & 1.57 \\
     3.3 & 8.1 & 13.5 & 1.45 & 1.55 \\
     4.6 & 8.18 & 15 & 1.45 & 1.51 \\
     6. & 8.2 & 15.7 & 1.52 & 1.51 \\
    \bottomrule
  \end{tabular}
  \label{tab:H12_Wscan_summary}
\end{table}
\begin{table}[h]
  \centering
  \caption{Global parameters for the scan in Ar pedestal concentration of the H12 scenario.}
  \begin{tabular}{>{\columncolor{green}}c c c c c c c c} 
    \toprule
    \rowcolor{green} 
     \textbf{$f_{Ar,top}$ $(10^{-3})$} & \textbf{$<Z_{eff}>$} & \textbf{$p_{top}$ $(kPa)$}& \textbf{$<f_{DT}>$} & \textbf{$\nu_{n_e}$} & \textbf{$P_{rad}$ $(MW)$} & \textbf{$Q$} & \textbf{$f_{LH,Schmidt}$}\\
    \midrule
     0.5 & 1.34 & 256 & 0.87 & 1.66 & 5.8 & 1.54 & 1.61 \\
     1.8 & 1.7 & 272 & 0.85 & 1.59 & 7.7 & 1.55 & 1.6 \\
     3.2 & 2.1 & 290 & 0.82 & 1.53 & 9.3 & 1.51 & 1.59 \\
     4.6 & 2.5 & 304 & 0.79 & 1.47 & 10.8 & 1.53 & 1.59 \\
     6 & 2.9 & 320 & 0.77 & 1.37 & 12 & 1.43 & 1.57 \\
     7.2 & 3.3 & 337 & 0.75 & 1.38 & 13.3 & 1.49 & 1.57 \\
     8.6 & 3.8 & 358 & 0.72 & 1.38 & 14.8 & 1.52 & 1.56 \\
    \bottomrule
  \end{tabular}
  \label{tab:H12_Arscan_summary}
\end{table}

\section{Impact of rotation on impurity transport}
\begin{figure}[h]
    \centering
    \includegraphics[width=0.45\linewidth]{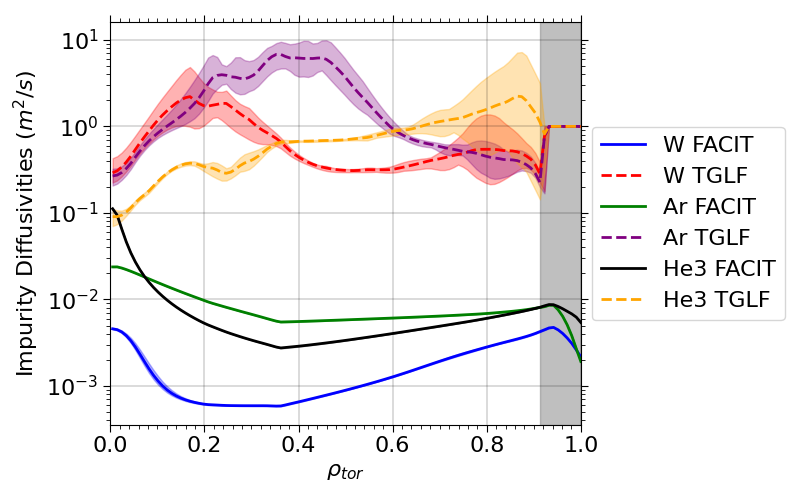}
    \includegraphics[width=0.45\linewidth]{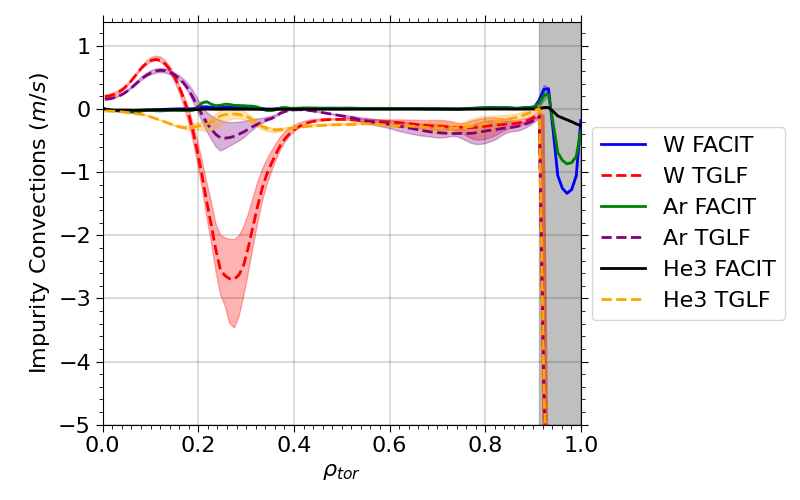}
    \caption{Scan in edge toroidal rotation for the H12 scenario. Profiles of turbulent (dashed) and neoclassical (solid) transport coefficients. On the left (right) the diffusivities (convections) are shown for all the impurities. The colored shades indicate the maximum and minimum values computed for the transport coefficients across the scan. The gray area is from top of pedestal to the separatrix.}
    \label{fig:vscan_profiles}
\end{figure}

\begin{figure}[h]
    \centering
    \includegraphics[width=0.45\linewidth]{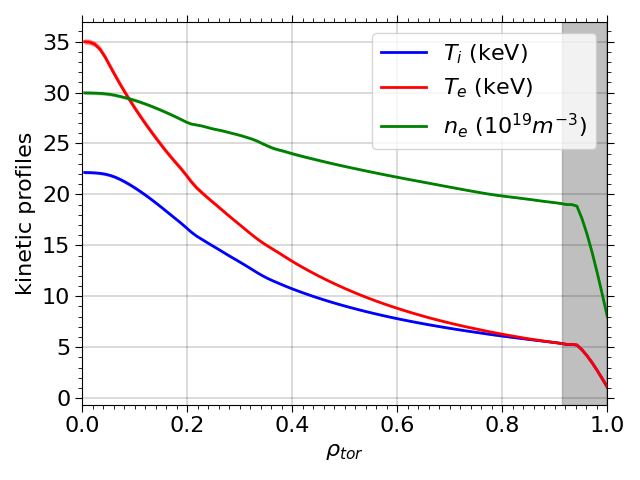}
    \includegraphics[width=0.45\linewidth]{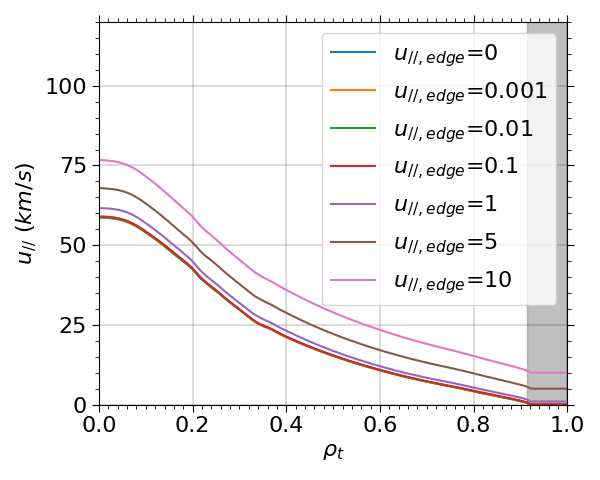}
    \caption{Kinetic (left) and parallel velocity (right) profiles for a scan in edge toroidal rotation of the H12 scenario. The colored shades indicate the maximum and minimum kinetic profiles computed across the scan. The gray area is from top of pedestal to the separatrix.}
    \label{fig:vscan_kinprofiles}
\end{figure}

\begin{figure}[h]
    \centering
    \includegraphics[width=0.45\linewidth]{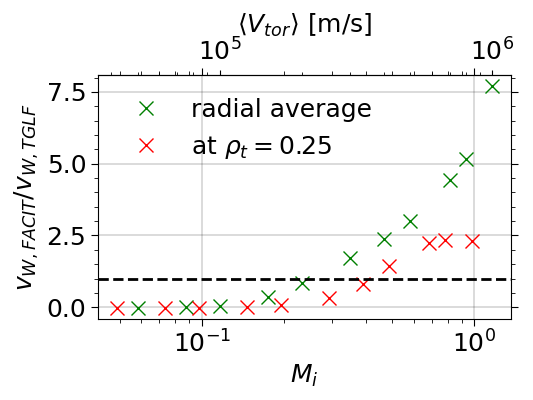}
    \includegraphics[width=0.45\linewidth]{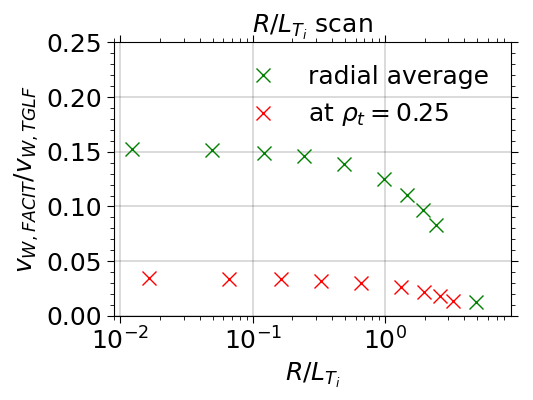}
    \includegraphics[width=0.45\linewidth]{vW_FACIT_avg_and_loc_gradTiscan.png}
    \includegraphics[width=0.45\linewidth]{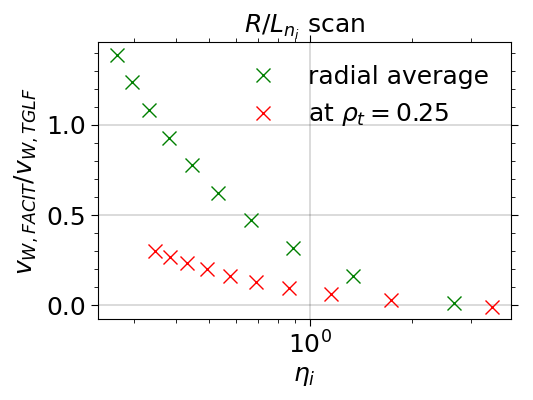}
    \caption{Ratios between W neoclassical and turbulent convection for different FACIT standalone scans. Top left: rotation (and Mach number) scan; top right: $R/L_{T_i}$ scan; bottom: $R/L_{n_i}$ scan. The bottom left (right) plot shows the convection ratio vs $R/L_{n_i}$ ($\eta_i$). Red indicates the value at $\rho_t=0.25$ and green shows the radial average up to the pedestal top.}
    \label{fig:FACIT_scans}
\end{figure}

The previously performed simulations assumed zero toroidal velocity. Although such an assumption is conservative in maximizing the turbulent transport \cite{waltz_theory_1998, kinsey_nonlinear_2005, camenen_impact_2009, barnes_turbulent_2011, howard_gyrokinetic_2021}, it is typically associated with low impurity accumulation \cite{angioni_neoclassical_2014, fajardo_analytical_2023}. Therefore, the analytical model developed in \cite{zimmermann_experimental_2024} has been implemented in ASTRA to predict core toroidal rotation and its effect on neoclassical and turbulent transport, respectively through the Mach number and radial electric field. This model consists of a set of analytical expressions aimed at predicting the momentum diffusivity, convection and intrinsic torque. It has been validated on AUG and D3D discharges \cite{zimmermann_analysis_2022, zimmermann_experimental_2024}. It is worth to mention that the model does not treat the effect of ripples \cite{fenzi_plasma_2011} and NTV torque \cite{park_self-consistent_2017, clement_neoclassical_2022}, which could lead to counter-current intrinsic torque in case of strong edge error fields. The description of such effects is beyond the scope of this paper. 
Simulations have been performed including momentum transport and using the analytical expressions up to the top of pedestal position. This choice is justified by the fact that the model was validated in the core. Therefore, a scan of flat toroidal rotation values from top of pedestal to the separatrix has been performed to account for uncertainties. This choice is supported by experimental evidence from Alcator C-Mod, which shows that the variation of the edge toroidal rotation in the absence of strong beam heating in an ICRH-dominated plasma is very limited and co-current \cite{mcdermott_edge_2009}. The resulting impurity transport coefficients are shown in figure \ref{fig:vscan_profiles}, while the kinetic and parallel velocity profiles are shown in figure \ref{fig:vscan_kinprofiles}.
The figures show respectively a small and no variation in the turbulent coefficients and kinetic profiles. Figure \ref{fig:vscan_profiles} highlights how the neoclassical transport is basically identical across the scan. This is justified by the low collisionality of the plasma, where the variation in rotation has a weak effect on neoclassical pinch and diffusivity \cite{fajardo_analytical_2023}. This observation has motivated a deeper analysis into the neoclassical transport coefficients obtained by FACIT. In fact, a multi-dimensional scan of the main ion temperature and density gradients and of the toroidal rotation have been performed around the nominal H12 simulation. Such scan aims at finding the thresholds above which the neoclassical transport becomes comparable with the anomalous one.
The ranges in which the parameters were varied are shown in table \ref{tab:FACIT_scan_ranges}.
\begin{table}[h]
  \centering
  \caption{Variation ranges of normalized $n_i$ and $T_i$ gradients and toroidal rotation for individual scans of FACIT standalone at $\rho_t=0.25$. The scaling factors refer to a nominal ASTRA+TGLF+FACIT simulation.}
  \begin{tabular}{>{\columncolor{brown}}c c c} 
    \toprule
    \rowcolor{brown} 
     \textbf{scan parameter} & \textbf{scaling factors} & \textbf{values} \\
    \midrule
     \textbf{$v_{tor}$} & -- & $[5 - 300]\times 10^4$ $[m/s]$ \\
     \textbf{$R/L_{T_i}$} & $[0.0025 - 1]$ & $[0.0125 - 5]$ \\
     \textbf{$R/L_{n_i}$} & $[1 - 10]$ & $[1.8 - 18]$ \\
    \bottomrule
  \end{tabular}
  \label{tab:FACIT_scan_ranges}
\end{table}
It is worth mentioning that the variation ranges of the scans have been intentionally chosen unrealistically wide as a numerical exercise to find the transition to neoclassically dominated impurity transport.
The results of the scans are summarized in figure \ref{fig:FACIT_scans}.
This figure shows several interesting trends which deserve further explanation. The toroidal velocity scan shows that the neoclassical transport becomes comparable to the turbulent one for Mach ion numbers above 0.25. Although such numbers can be reached in tokamak plasmas, the figure shows that for this SPARC scenario this would correspond to toroidal velocities bigger than 300 $km/s$, which is an order of magnitude higher than the maximum values found in the momentum transport simulations shown earlier in this section. The normalized ion temperature gradient scan shows that even approaching near-null values of $R/L_{T_i}$ the ratio between neoclassical and anomalous pinch asymptotically tends to a maximum value of 0.15. Finally, the density gradient scan, shown at the bottom left of the figure, illustrates that the average $v_{W,NC}/v_{W,turb}$ reaches $\sim$ 1 only for unrealistic values of $R/L_{n_i}$ which are an order of magnitude higher than those found in nominal ASTRA flux-matched simulations. For comparison, also $\eta_i=L_{n_i}/L_{T_i}$ is plotted at the bottom right of the figure. The scan highlights how $\eta_i<0.35$ would be needed to obtain $v_{W,NC}\sim v_{W,turb}$.
The results presented in this section demonstrate the robust predominance of anomalous impurity transport, indicating that only unrealistically high toroidal rotation or normalized density gradient values—unlikely to be achieved in reactor-relevant devices—would make neoclassical transport comparable in magnitude.

\section{Sensitivity of fusion and impurity peaking to DT fuel mix composition}
\begin{figure}[h]
    \centering
    \includegraphics[width=0.45\linewidth]{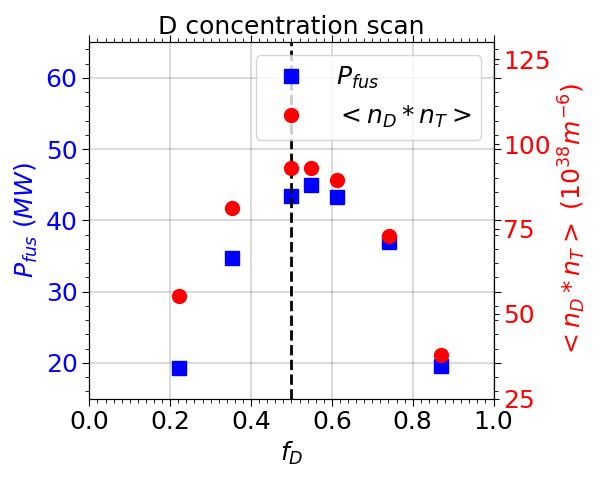}
    \includegraphics[width=0.45\linewidth]{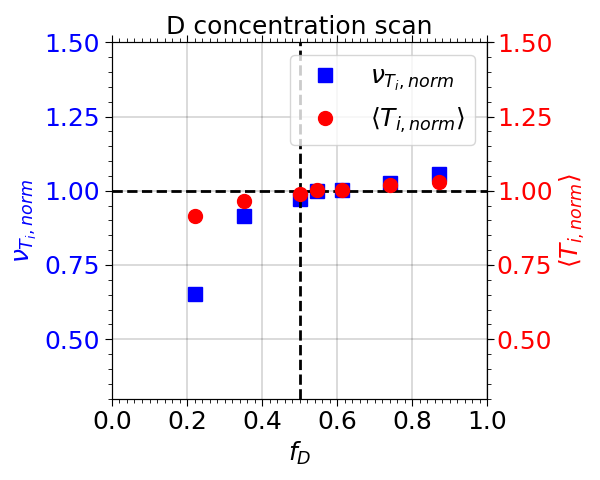}
    \caption{Left: fusion power (blue) and $n_Dn_T$ (red) for an H12 scan of D pedestal concentration. Right: average ion temperature (red) and $T_i$ peaking (blue), normalized to the nominal ASTRA case (i.e. 50-50\% DT)). For both plots, the X-axis indicates the D concentration at top of pedestal with respect to the total D+T density.}
    \label{fig:Tscan_Pfus}
\end{figure}

\begin{figure}[h]
    \centering
    \includegraphics[width=0.45\linewidth]{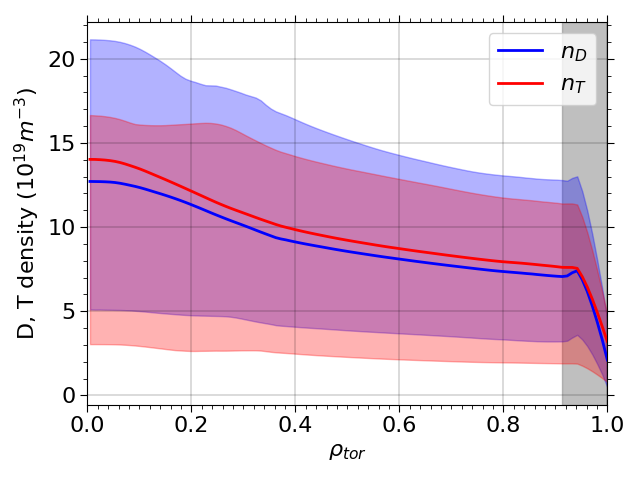}
    \includegraphics[width=0.45\linewidth]{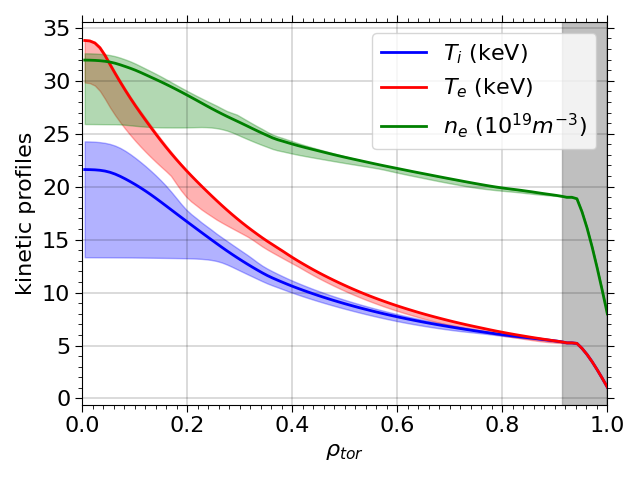}
    \caption{On the left the D (blue) and T (red) density profiles are shown, while on the right the kinetic profiles are pictured. The colored shades indicate the maximum and minimum values computed for the profiles. The gray area is from top of pedestal to the separatrix.}
    \label{fig:Tscan_kinprofiles}
\end{figure}

In the previous simulations, the main ion species has been assumed with a 50-50\% DT fuel composition, lumping the D and T masses into a unique species with mass 2.5. However, a different fuel mixture leads to different fusion power, which in turn affects the kinetic profiles. This could indirectly impact the impurity transport, which may provide a different peaking level via diverse turbulent / neoclassical contributions. This argument has motivated additional simulations, performed scanning the D concentration at the top of pedestal and evolving it separately from T as an additional species in ASTRA. Therefore, in the new simulations, electrons, T, He3 and Ar are simulated, D is imposed by quasi-neutrality, and W is not included, due to the limits of 5 species imposed by the present implementation of TGLF in ASTRA. W has been excluded from the simulations because it is the most diluted species in the plasma and Ar is needed to maintain the target $Z_{eff}$ value. The D pedestal concentration (i.e. $f_D$, with respect to the total D+T density) has been scanned from 0.22 to 0.88. The results are shown in figure \ref{fig:Tscan_Pfus}, while figure \ref{fig:Tscan_kinprofiles} shows the D and T densities, together with the kinetic profiles.
\begin{figure}
    \centering
    \includegraphics[width=0.45\linewidth]{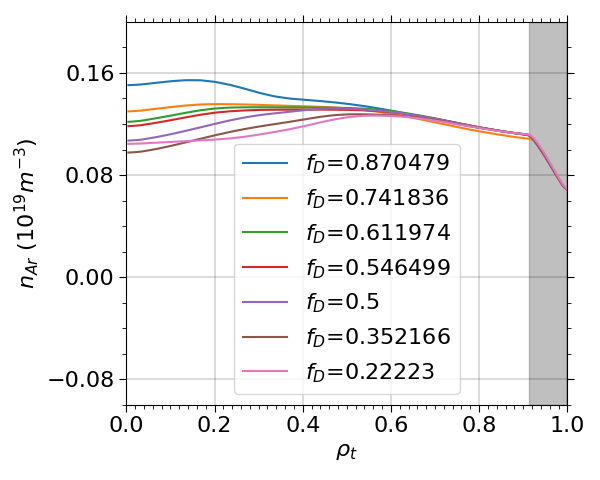}
    \includegraphics[width=0.45\linewidth]{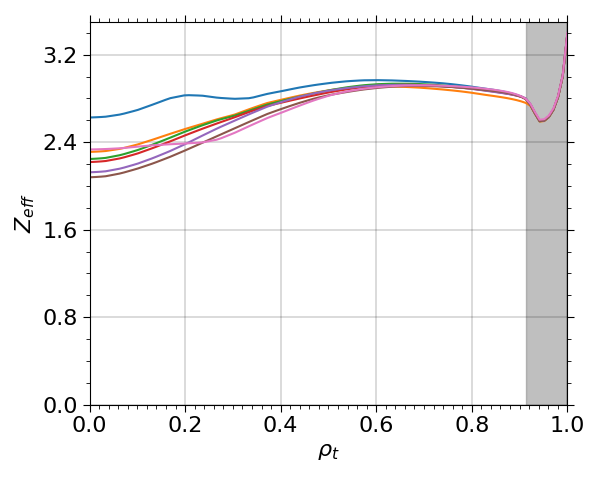}
    \includegraphics[width=0.45\linewidth]{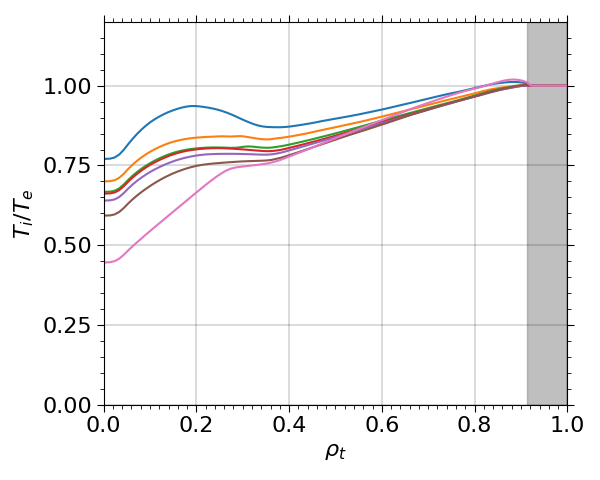}
    \includegraphics[width=0.45\linewidth]{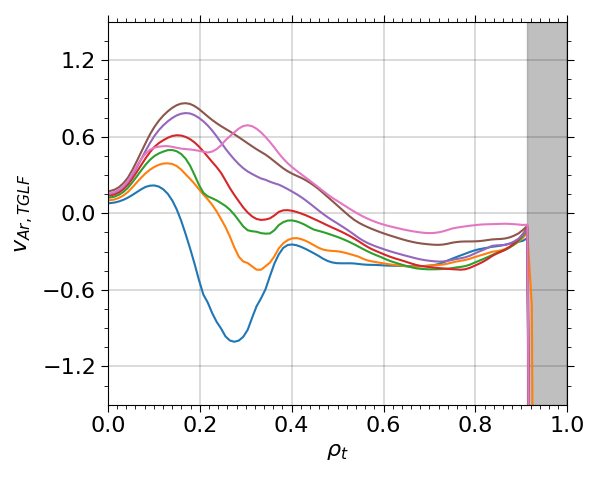}
    \caption{Profiles for a scan in D concentration of the H12 scenario. Top left: Ar density; top right: effective charge; bottom left: ion to electron temperature ratio; bottom right: anomalous Ar convection. The gray area is from top of pedestal to the separatrix.}
    \label{fig:Tscan_profiles}
\end{figure}
\begin{figure}[h]
    \centering
    \includegraphics[width=0.45\linewidth]{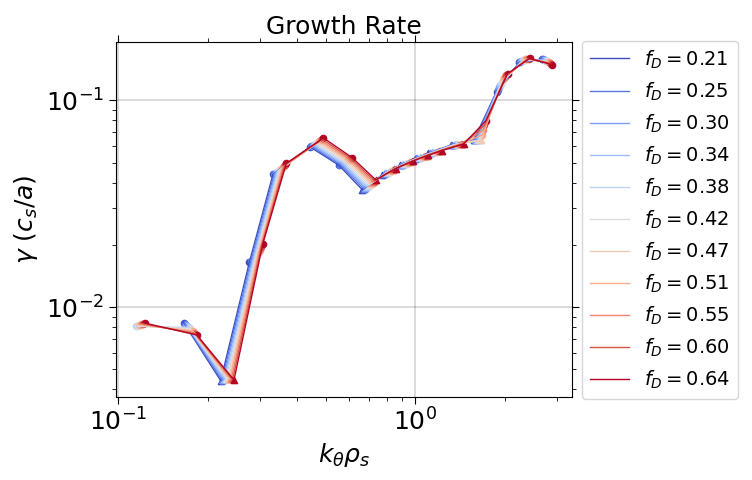}
    \includegraphics[width=0.45\linewidth]{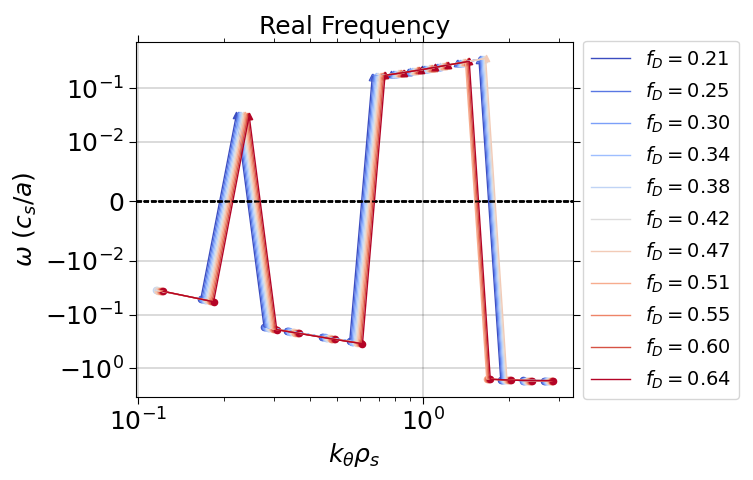}
    \includegraphics[width=0.45\linewidth]{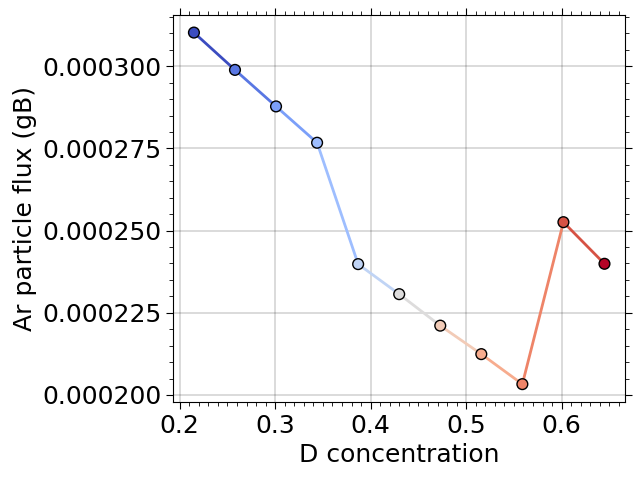}
    \includegraphics[width=0.45\linewidth]{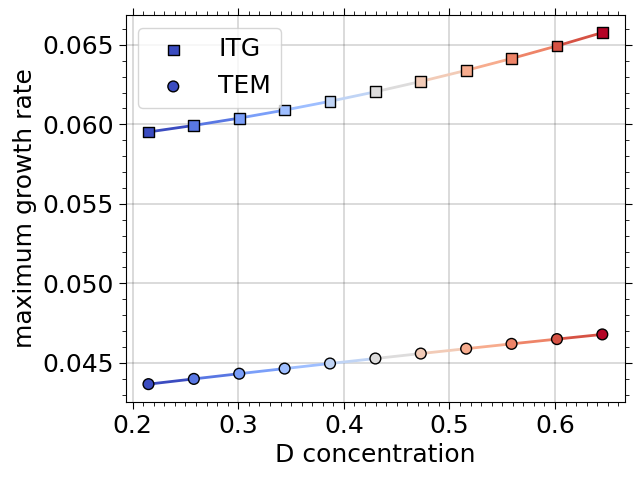}
    \caption{Turbulent spectra and global quantities of TGLF standalone simulations scanning the D concentration ($f_D$) and modifying the T density to enforce quasi-neutrality. Top left: growth rates; top right: mode frequencies; bottom left: Ar particle flux; bottom right: most unstable ITG and TEM growth rates. Different colors are for different $f_D$ values.}
    \label{fig:Tscan_TGLF}
\end{figure}
Figure \ref{fig:Tscan_Pfus} indicates that the maximum fusion power is around 55-45\% DT fuel mix. This result is unexpected and requires further investigation. Interestingly, the dependence of fusion power on $f_D$ is not symmetric with respect to the 50-50\% composition. In particular, higher D concentrations give higher fusion power at similar $n_Dn_T$ values. Since $P_{fus}$ scales as $n_Dn_TT_i^{C}$, where $C>0$, this indicates that the DT fuel composition affects the average ion temperature profile, predicting higher values at higher $f_D$. In order to verify this, the average $T_i$ and its peaking are shown in the right plot of figure \ref{fig:Tscan_Pfus}. The higher ion temperature at higher D concentrations explains why the maximum fusion power is found at 55-45\% DT fuel composition. The variation of the $T_i$ profile can be seen in the right plot of figure \ref{fig:Tscan_kinprofiles}. Figure \ref{fig:Tscan_profiles} shows that the Ar density in the inner core increases with $f_D$, leading to higher $Z_{eff}$, which in turn affect the ITG stability, setting different $T_i/T_e$ values.
The effect of increased radiation at higher Ar densities on the electron temperature is negligible, due to its small relative variation and the stiffness of the $T_e$ profile. The difference in the Ar peaking is related to the effect that the main ion mass has on the background turbulence (i.e. isotope effect). In fact, it has been observed that D has a higher turbulent growth rate compared to T, with fixed other conditions \cite{belli_asymmetry_2021, snoep_characterization_2025}. This typically implies a higher heat flux, while its effect on particle fluxes is non-trivial and it depends on the turbulent regime \cite{angioni_direction_2006, angioni_particle_2009, fable_role_2010}. The bottom right plot of figure \ref{fig:Tscan_profiles} shows higher impurity inward convection at higher $f_D$. A higher D concentration in the fuel mix can be considered similar to a DT lumped species with lower mass. The observed isotope effect on particle transport has been confirmed by standalone TGLF simulations, performed scanning the D concentration and enforcing the T density with quasi-neutrality. The results, shown in figure \ref{fig:Tscan_TGLF}, show an ITG-dominated regime, where the Ar outward flux decreases with $f_D$. The same trend has been observed for the electron particle flux.
An additional sensitivity study, varying the Ar density gradient by $\pm50\%$ has shown an Ar flux variation $<10\%$, proving the robustness of the results. Since the regime is ITG dominated and the impurity flux decreases at higher linear growth rates, the pure pinch should prevail on thermo-diffusion \cite{angioni_direction_2006}. These results suggest that if a 55-45\% DT fuel mix is not available, higher D concentrations are preferable and can still provide reasonable fusion power. For example a 75-25\% DT mix shows 20\% higher fusion power compared to 25-75\% composition. However, this study does not aim at describing the physics of isotope effect in detail, therefore a more rigorous analysis, e.g. comparing with non linear gyrokinetics and separating the convection in pure pinch, thermo- and roto-diffusion, is beyond the scope of the paper.

\section{Conclusions}
In this paper, the impurity transport for three SPARC H-modes has been investigated. The results have shown that the turbulence dominates over the neoclassical contribution for Ar and W impurities. A successful benchmark with previously published simulations has been provided, highlighting that the assumption of flat impurity concentrations is a reasonable approach to quantify fusion performance variations in extensive databases. Such evidence has been shown at different values of ICRH input power and top of pedestal densities. The modeling used in this paper assumed an ad-hoc edge transport to set impurity concentrations as boundary conditions at the top of pedestal. This choice is imposed by the lack of accurate turbulent model in stiff H-mode pedestals, but the precise value of pedestal concentrations is uncertain. Therefore, the low current H-mode scenario has been selected to perform scans of Ar and W source, which determined different boundary conditions at pedestal, covering a wide and conservative range. The results showed a minimal impact of W concentration variations on the impurity profiles. On the other hand, the Ar source scan has shown an interesting non-trivial interplay between different effects, including pedestal pressure, density peaking and DT dilution, which counterbalanced each other resulting in the plasma performance being insensitive to the Ar concentration and the associated $Z_{eff}$. Most of the performed simulations assumed null rotation. Without external torque sources, this is a reasonable assumption for SPARC, which will operate without Neutral Beam Injection. However, rotation has been observed to have an impact on both turbulent and neoclassical transport. Therefore, a set of ASTRA simulations including a reduced model for core rotation has been used, varying a constant edge toroidal velocity in a wide range as boundary condition for core calculations. The results have shown minimal impact on the electron, main ion and impurities profiles. In order to assess the robustness of these predictions, additional scans of FACIT standalone have been performed, to determine variables ranges in which the neoclassical impurity transport becomes competitive with the turbulent one. The variation of normalized $T_i$ gradient has shown that even assuming flat profile will give a maximum contribution of neoclassical transport of about 15\%. However, high fusion conditions should lead to significant power flow and relatively peaked temperature profiles. The scan in normalized $n_i$ gradient has shown that the neoclassical transport becomes comparable with the turbulent one only at $R/L_{n_e}>12$, which is unrealistic. The scan in toroidal velocity revealed that the transport coefficients calculated by FACIT reach TGLF values only above $200 km/s$ (i.e. Mach number 0.2), which is one order of magnitude higher than the maximum value found in ASTRA. Such a high rotation seems unrealistic for the analyzed scenario. All the simulations performed in the paper, including the latest sensitivity scans, showed that the turbulent W transport is robustly higher than the neoclassical component, across multiple variations of uncertain parameters. These findings align with ITER predictions from existing literature \cite{fajardo_theory-based_2025}.
Finally, an additional study has been performed to study the effect of assuming different DT fuel mix compositions on the impurity transport and plasma performance. Surprisingly, the maximum fusion power has been found at 55-45\% DT fuel mix composition, with high D concentrations preferable with respect to symmetrically high T concentrations. This unexpected result comes from the isotope effect on impurity transport. In particular, higher D concentrations showed higher linear growth rates, which in the ITG dominated regime studied here increased the Ar pinch, leading to higher $Z_{eff}$ and ion temperature peaking.

\newpage
\section*{Acknowledgments}
The authors thank the MIT PSFC for its constructive feedback, in particular the MFE Integrated Modeling group, John Wright for technical support about computational resources and Conor Perks for interesting discussions about impurity transport. The authors thank Teobaldo Luda, Michael Bergmann and Emiliano Fable for sharing their experiences with the ASTRA code.
The authors acknowledge the use of ChatGPT during the article editing phase. This research used resources of the National Energy Research Scientific Computing Center, a DOE Office of Science User Facility using NERSC award FES-ERCAP0032161, for the EPED simulations used to train the neural nework model. The ASTRA simulations (ASTRA from main branch with hash a00f496a5489e12bbdbdc02dc38482057dd43b0b) presented in this paper were performed on the MIT-PSFC partition of the Engaging cluster at the MGHPCC facility (www.mghpcc.org) which was funded by DoE grant number DE-FG02-91-ER54109. \\
\textit{This work was supported by CFS under RPP020 fundings}.\\

\printbibliography

\end{document}